\documentclass[11pt]{article}
\usepackage{amsmath,amssymb,graphicx}
\usepackage{hyperref}
\usepackage[nosort]{cite}
\usepackage{multicol,longtable}
\usepackage{tensor}
\usepackage{multirow}
\usepackage{nicefrac}
\usepackage{cases}
\usepackage[vcentermath]{youngtab}
\usepackage{blindtext}
\usepackage{float}
\usepackage{rotating}

\usepackage{cite}
\usepackage{amsmath,amsfonts,amssymb,calrsfs}
\usepackage[small,bf,hang]{caption}
\usepackage{slashed}
\usepackage{subfig}
\usepackage{latexsym,epsfig}

\usepackage[final]{showkeys}
\usepackage{amsfonts}
\usepackage{mathrsfs}
\usepackage{dsfont}
\usepackage[Symbol]{upgreek}

\usepackage{tcolorbox}
\usepackage{empheq}

\usepackage{ytableau}
\usepackage{titlesec}

\titleformat{\paragraph}
{\normalfont\normalsize\bfseries}{\theparagraph}{1em}{}
\titlespacing*{\paragraph}
{0pt}{3.25ex plus 1ex minus .2ex}{1.5ex plus .2ex}

\DeclareMathAlphabet{\matcal}{OMS}{cmsy}{m}{n}

\DeclareMathAlphabet{\mathpzc}{OT1}{pzc}{m}{it}

\newcommand{\uA}{{\underline{A}}}
\newcommand{\uB}{{\underline{B}}}
\newcommand{\uC}{{\underline{C}}}
\newcommand{\uD}{{\underline{D}}}
\newcommand{\uE}{{\underline{E}}}
\newcommand{\uF}{{\underline{F}}}

\newcommand{\uM}{{\underline{M}}}
\newcommand{\uN}{{\underline{N}}}
\newcommand{\uP}{{\underline{P}}}

\newcommand{\IU}{(U^{-1})}
\newcommand{\gU}{{\cal U}}

\newcommand{\gM}{{\cal M}}
\newcommand{\gL}{{\cal L}}
\newcommand{\cA}{{\cal A}}
\newcommand{\cB}{{\cal B}}
\newcommand{\VI}{({\cal V}^{-1})}

\newcommand{\SU}[1]{\mathrm{SU}(#1)}
\newcommand{\SL}[1]{\mathrm{SL}(#1)}

\newcommand{\SO}[1]{\mathrm{SO}(#1)}

\newcommand{\be}{\begin{align}}
\newcommand{\ee}{\end{align}}

\makeatletter

\@addtoreset{equation}{section}
\makeatother

\newcommand{\EOn}[2]{\left(#1\right)_{#2}}

\newcommand{\bea}{\begin{eqnarray}}
\newcommand{\eea}{\end{eqnarray}}

\def\*{\partial}

\def\={\!=\!}

\usepackage{cite}
\usepackage{amsmath,amsfonts,amssymb}

\usepackage{color}
\definecolor{red}{rgb}{1,0,0}
\definecolor{lred}{rgb}{0.3,0,0}
\definecolor{green}{rgb}{0,0.6,0}
\definecolor{blue}{rgb}{0,0,1}
\definecolor{violet}{rgb}{0.8,0,0.8}

\definecolor{darkred}{rgb}{0.65,0.15,0}
\definecolor{darkgreen}{rgb}{.05,.5,.05}
\hypersetup{pdfborder={0 0 0},colorlinks=true,urlcolor=darkred,citecolor=blue,linkcolor=darkred,linktocpage=true}

\newdimen\squaresize \squaresize=12pt
\newdimen\thickness \thickness=0.7pt

\def\square#1{\hbox{\vrule width \thickness
   \vbox to \squaresize{\hrule height \thickness\vss
      \hbox to \squaresize{\hss#1\hss}
   \vss\hrule height\thickness}
\unskip\vrule width \thickness} \kern-\thickness}

\def\cut#1{\hbox{\vrule width-1 \thickness
   \vbox to \squaresize{\hrule height-1 \thickness\vss
      \hbox to \squaresize{\hss#1\hss}
   \vss\hrule height-1\thickness}
\unskip\vrule width +4 \thickness} \kern-\thickness}

\def\vsquare#1{\vbox{\square{$#1$}}\kern-\thickness}

\begin{document}

\begin{titlepage}

\vfill
\begin{flushright}
HU-EP-20/24
\end{flushright}

\vfill

\begin{center}

   {\LARGE \bf  
   Kaluza-Klein Spectrometry\\[1ex] 
   from Exceptional Field Theory
   }
   \vskip 2cm

{\large\bf {Emanuel Malek${\,}^a$\footnote{\tt emanuel.malek@physik.hu-berlin.de}, Henning Samtleben${\,}^b$\footnote{\tt henning.samtleben@ens-lyon.fr}}}
\vskip .6cm
{\it ${}^a$
	Institut f\"{u}r Physik, Humboldt-Universit\"{a}t zu Berlin,\\
	IRIS Geb\"{a}ude, Zum Gro{\ss}en Windkanal 6, 12489 Berlin, Germany}\\ \ \\
\vskip .2cm

{\it ${}^b$
Univ Lyon, Ens de Lyon, Univ Claude Bernard, CNRS,\\
Laboratoire de Physique, F-69342 Lyon, France} \\ \ \\
\vskip .2cm

\end{center}

\vfill

\begin{center} 
\textbf{Abstract}

\end{center} 
\begin{quote}

Exceptional field theories yield duality-covariant formulations of higher-dimensional supergravity. They have proven to be an efficient tool for the construction of consistent truncations around various background geometries. In this paper, we demonstrate how the formalism can moreover be turned into a powerful tool for computing the Kaluza-Klein mass spectra around these backgrounds. Most of these geometries have little to no remaining symmetries and their spectra are accessible to standard methods only in selected subsectors. The present formalism not only grants access to the full Kaluza-Klein spectra but also provides the scheme to identify the resulting mass eigenstates in higher dimensions. As a first illustration, we rederive in compact form the mass spectrum of IIB supergravity on $S^5$. We further discuss the application of our formalism to determine the mass spectra of higher Kaluza-Klein multiplets around the warped geometries corresponding to some prominent ${\cal N}=2$ and ${\cal N}=0$ AdS vacua in maximal supergravity. 

\end{quote} 
\vfill
\setcounter{footnote}{0}
\end{titlepage}

\clearpage
\setcounter{page}{1}

\tableofcontents

\section{Introduction}

Whenever a higher-dimensional theory is compactified, towers of infinitely many massive fields arise in the lower-dimensional theory. These Kaluza-Klein towers are the lower-dimensional signature of the compactification space and often play a crucial role in the compactified theory. For example, in phenomenological models arising out of string theory, these Kaluza-Klein towers would correspond to massive particles but may also indicate potential instabilities of the background. On the other hand, in the AdS/CFT correspondence, the masses of the Kaluza-Klein towers are mapped to the conformal dimensions of operators in strongly-coupled CFTs, that cannot be computed directly except for protected operators. Despite the universality and importance of Kaluza-Klein towers, calculating their masses is an exceedingly difficult undertaking. Indeed, obtaining the Kaluza-Klein spectrum of supergravity compactifications has hitherto only been possible for coset spaces, while on general backgrounds this has only been achieved for the spin-2 towers.

This paper is a detailed account of the results of \cite{Malek:2019eaz}. There we announced a new method based on Exceptional Field Theory (ExFT), which allows us to compute the full Kaluza-Klein spectrum for any vacuum of a maximal gauged supergravity arising from a consistent truncation of 10- or 11-dimensional supergravity. This includes vacua with few or no (super-)symmetries, whose Kaluza-Klein spectra were previously inaccessible. ExFT is a duality-covariant reformulation of maximal 10-/11-dimensional supergravity, which unifies fluxes and gravitational degrees of freedom. Since the Kaluza-Klein fluctuations mix between the flux and gravitational sectors of supergravity, this makes ExFT a natural formulation within which to study this problem.

Indeed, as we develop here, we can build on the efficient ExFT description of consistent truncations to maximal gauged supergravity \cite{Aldazabal:2011nj,Geissbuhler:2011mx,Grana:2012rr,Berman:2012uy,Lee:2014mla,Hohm:2014qga} to obtain a remarkably simple expression for the Kaluza-Klein fluctuations around any vacuum of the lower-dimensional gauged supergravity. The fluctuation Ansatz takes the form of the lower-dimensional supergravity multiplet, making up the consistent truncation, tensored with the scalar harmonics of the maximally symmetric point of the lower-dimensional supergravity. As a result, the Ansatz is \emph{non-linear} in the fields of the lower-dimensional supergravity multiplet. Due to this non-linearity, it is straightforward to compute the Kaluza-Klein spectrum for \emph{any} vacuum of the lower-dimensional supergravity arising from the consistent truncation.
\medskip

There are several benefits to our approach:
\begin{itemize}
	\item The fluctuations of all supergravity fields are parametrised in terms of a common set of ``scalar harmonics''. In contrast, in the traditional approach, fields in different Lorentz representations require different harmonics.
	\item The scalar harmonics are computed at the maximally symmetric point of the lower-dimensional supergravity, even if we are interested in another vacuum of the lower-dimensional supergravity with a much smaller symmetry group.
	\item As a consequence, we can, for the first time, compute the Kaluza-Klein spectrum around vacua with few or no (super-)symmetries, including non-supersymmetric vacua, such as the prominent non-supersymmetric $\SO{3} \times \SO{3}$ AdS$_4$ vacuum of 11-dimensional supergravity \cite{Godazgar:2014eza}.
	\item The states of every BPS multiplet live in the same Kaluza-Klein level, making the identification of supermultiplets in 10/11 dimensions considerably easier than using the traditional approach, where BPS multiplets are scattered amongst different Kaluza-Klein levels.
	\item Using the dictionary between the ExFT and the original supergravity variables, it is straightforward to identify
	the higher-dimensional origin of the resulting mass eigenstates.
\end{itemize}

The paper is structured as follows. We begin with a review of the relevant aspects of ExFT in section \ref{s:Review}. In section \ref{s:Fluctuation}, we then describe how to efficiently parametrise the Kaluza-Klein fluctuations in ExFT. In section \ref{sec:mass}, we show how this leads to compact expressions for the mass matrices of the Kaluza-Klein towers, including the vector and the scalar fields. We next demonstrate the power of the formalism by applying it to several prominent AdS vacua of 10- and 11-dimensional supergravity in section \ref{s:Examples}. In particular, we show how our approach leads to a very efficient computation of the Kaluza-Klein spectrum of AdS$_5 \times S^5$ and the identification of the mass eigenstates within IIB supergravity.
We then elaborate on the results announced in \cite{Malek:2019eaz}, by
\begin{itemize}
	\item computing the spectrum of the first Kaluza-Klein level above the ${\cal N}=8$ supergravity of the $\SU{2} \times \mathrm{U}(1)$-invariant AdS$_5$ vacuum of IIB supergravity \cite{Pilch:2000ej} that is dual to the Leigh-Strassler CFT \cite{Leigh:1995ep},
	\item giving the full bosonic Kaluza-Klein spectrum of the $\SU{3} \times \mathrm{U}(1)$-invariant AdS$_4$ vacuum of 11-dimensional supergravity \cite{Corrado:2001nv}, dual to a quadratic deformation of ABJM,
	\item reviewing the computation of \cite{Malek:2020mlk} of the Kaluza-Klein spectrum of the non-supersymmetric $\SO{3} \times \SO{3}$-invariant AdS$_4$ vacuum of 11-dimensional supergravity, and the appearance of tachyonic scalars at higher Kaluza-Klein levels.
\end{itemize}
Finally, we conclude with a summary of our results and outlook on further problems to be tackled in section~\ref{s:Conclusions}.

\section{Exceptional field theory} \label{s:Review}

In this section, we briefly review the structure of the relevant exceptional field theories (ExFTs), based on the exceptional groups $\rm{E}_{6(6)}$ and $\rm{E}_{7(7)}$, respectively. We refer to~\cite{Hohm:2013pua,Hohm:2013vpa,Hohm:2013uia,Baguet:2015xha} for further details.
These are the duality-covariant formulations of maximal supergravity in ten and eleven dimensions,
tailored to describe compactifications to $D=5$ and $D=4$ dimensions, respectively.

\subsection{E$_{6(6)}$ ExFT}

The Lagrangian of E$_{6(6)}$ exceptional field theory (ExFT) is modelled after maximal five-dimensional supergravity~\cite{Cremmer:1980gs,deWit:2004nw}.
Its bosonic field content is given by
\begin{equation}
\left\{
g_{\mu\nu}, {\cal M}_{MN}, {\cal A}_{\mu}{}^M, {\cal B}_{\mu\nu\,M} \right\}
\,,\qquad
\mu, \nu =0, \dots, 4\;,\quad
M=1, \dots, 27\,,
\label{ExFTfieldsE6}
\end{equation}
and combines a $5\times 5$ `external' metric $g_{\mu\nu}$ with an `internal' $27\times27$ generalised metric ${\cal M}_{MN}$, the latter parametrizing the coset space ${\rm E}_{6(6)}/{\rm USp}(8)$\,. Therefore, the generalised metric can be expressed in terms of a generalised vielbein
\begin{equation} \label{eq:GenMetricE6}
 \gM_{MN} = {\cal E}_M{}^{\uM}\, {\cal E}_N{}^{\uN}\, \delta_{\uM\uN} \,,
\end{equation}
where the generalised vielbein, ${\cal E}_M{}^{\uM}$, is an ${\rm E}_{6(6)}$-valued matrix. Vector and tensor fields ${\cal A}_{\mu}{}^M$ and ${\cal B}_{\mu\nu\,M}$ are labelled by an index $M$ in the (anti-)fundamental representation of ${\rm E}_{6(6)}$\,. These are the fields of maximal five-dimensional supergravity; however, all of them are still living on the full higher-dimensional spacetime.
The complete bosonic Lagrangian reads
\begin{equation}
 {\cal L}_{\rm ExFT6} = \sqrt{|g|}\, \Big( \widehat{R}
 +\frac{1}{24}\,g^{\mu\nu}{\cal D}_{\mu}{\cal M}^{MN}\,{\cal D}_{\nu}{\cal M}_{MN}
-\frac{1}{4}\,{\cal M}_{MN}{\cal F}^{\mu\nu M}{\cal F}_{\mu\nu}{}^N
 +\sqrt{|g|}{}^{-1}{\cal L}_{\rm top}
-V(g,{\cal M})\Big) \,.
\label{ExFTE6}
\end{equation}
It is invariant under generalised internal diffeomorphisms whose action on the scalar matrix ${\cal M}_{MN}$ has
the generic form \cite{Coimbra:2011ky,Berman:2012vc}
\begin{equation}
\delta_\Lambda {\cal M}_{MN}=
{\cal L}_\Lambda {\cal M}_{MN} = \Lambda^K \partial_K {\cal M}_{MN} 
+2\,\alpha_d\,\partial_L\Lambda^K\,\mathbb{P}^K{}_L{}^P{}_{(M}\,{\cal M}_{N)P}
\,.
\label{gen_diff_M}
\end{equation}
Here, $\mathbb{P}^K{}_L{}^P{}_{M}$ is the projector on the adjoint representation of the duality group E$_{d(d)}$,
which for ${\rm E}_{6(6)}$ takes the explicit form
\begin{equation}
\mathbb{P}^M{}_N{}^K{}_L
=
\frac1{18}\,\delta_N{}^M\delta_L{}^K + \frac16\,\delta_N{}^K\delta_L{}^M
-\frac53\,d_{NLR}d^{MKR}\,, 
\label{PadjE6}
\end{equation}
in terms of the totally symmetric cubic ${\rm E}_{6(6)}$-invariant tensor $d_{KMN}$\,.
The constant $\alpha_d$ in (\ref{gen_diff_M}) is determined by closure of the diffeomorphism algebra
and is equal to $\alpha_6=6$ for ${\rm E}_{6(6)}$\,.
The scalar fields in the Lagrangian (\ref{ExFTE6}) couple via a gauged sigma model on the coset space ${\rm E}_{6(6)}/{\rm USp}(8)$\,. Accordingly, ${\cal M}^{MN}$ denotes the matrix inverse to ${\cal M}_{MN}$, and the covariant derivatives are defined as
\begin{equation}
{\cal D}_{\mu}{\cal M}_{MN} =
(\partial_\mu -{\cal L}_{{\cal A}_\mu})\,{\cal M}_{MN}
\,,
\label{covDE6}
\end{equation}
corresponding to the action of (\ref{gen_diff_M}).

The Einstein-Hilbert term is constructed from the modified Ricci scalar $\widehat{R}$, constructed
from the external metric $g_{\mu\nu}$ in the standard way upon covariantising derivatives under internal diffeomorphisms
$\partial_\mu g_{\nu\rho} \rightarrow \partial_\mu g_{\nu\rho} - {\cal A}_\mu{}^K \partial_K g_{\nu\rho}$\,.
The non-Abelian field strengths in (\ref{ExFTE6}) are given by
\begin{equation}
{\cal F}_{\mu\nu}{}^N  =   2\, \partial_{[\mu} {\cal A}_{\nu]}{}^N 
-2\,{\cal A}_{[\mu}{}^K \partial_K {\cal A}_{\nu]}{}^N 
+10\, d^{NKR}d_{PLR}\,{\cal A}_{[\mu}{}^P\,\partial_K {\cal A}_{\nu]}{}^L
+ 10 \, d^{NKL}\,\partial_K {\cal B}_{\mu\nu\,L}
\,,
\end{equation}
with a St\"uckelberg-type coupling to the two-form tensors ${\cal B}_{\mu\nu\,N}$. In turn, the topological term ${\cal L}_{\rm top} $
is defined via its derivative
\begin{equation}
d{\cal L}_{\rm top} \propto 
d_{MNK}\,{\cal F}^M \wedge  {\cal F}^N \wedge  {\cal F}^K
-40\, d^{MNK}{\cal H}_M\,  \wedge \partial_N{\cal H}_K
\,,
\end{equation}
in terms of the field strengths ${\cal F}_{\mu\nu}{}^M$ and ${\cal H}_{\mu\nu\rho\,M} =
3\,{\cal D}_{[\mu} {\cal B}_{\nu\rho]\,M} + \dots$, with the ellipses denoting Chern-Simons type couplings
whose explicit form will not be relevant for this paper.
Finally, the potential term $V(g,{\cal M})$ in (\ref{ExFTE6}) is built from bilinears in internal derivatives and reads
\begin{equation}\label{fullpotential}
 \begin{split}
  V(g,{\cal M}) \ = \ &-\frac{1}{4\,\alpha_d}{\cal M}^{MN}\partial_M{\cal M}^{KL}\,\partial_N{\cal M}_{KL}
  +\frac{1}{2} {\cal M}^{MN}\partial_M{\cal M}^{KL}\partial_L{\cal M}_{NK}\\
  &-\frac{1}{2}\,g^{-1}\partial_Mg\,\partial_N{\cal M}^{MN}-\frac{1}{4}  {\cal M}^{MN}g^{-1}\partial_Mg\,g^{-1}\partial_Ng
  -\frac{1}{4}{\cal M}^{MN}\partial_Mg^{\mu\nu}\partial_N g_{\mu\nu}\;. 
 \end{split} 
\end{equation}

In the formulation (\ref{ExFTE6}), the internal coordinates are embedded into the 
27-dimensional representation of ${\rm E}_{6(6)}$ with derivatives denoted as $\partial_M$.
Gauge invariance of the action requires the so-called section constraint, expressed as 
a condition bilinear in internal derivatives
\begin{equation}
d^{KMN}\,\partial_M \Phi_1 \partial_N \Phi_2  = 0
\,,
\label{sectionE6}
\end{equation}
for any couple of fields $\{\Phi_1, \Phi_2\}$\,.
The section constraint (\ref{sectionE6}) can be solved by breaking E$_{6(6)}$ according to
\begin{equation}
	\begin{split}
\rm{E}_{6(6)} \supset~ {\rm SL}(6) \times {\rm SL}(2) &~\supset~  {\rm SL}(6) \times {\rm GL}(1)_{{\rm 11}} 
\,,\\
{\bf 27} \longrightarrow (6,2) + (15,1) &\longrightarrow  6_{+1} + 15'_{0} + 6_{-1}
\,, 
\label{solvesectionE6A}
\end{split}
\end{equation}
and restricting the coordinate dependence of all fields to the first six coordinates. Upon this choice, the Lagrangian
(\ref{ExFTE6}) becomes equivalent to full eleven-dimensional supergravity. In turn, type IIB supergravity is recovered
upon choosing a second inequivalent solution of the section constraint based on the group decomposition
\begin{equation}
	\begin{split}
\rm{E}_{6(6)} &~\supset~ {\rm SL}(5)\times{\rm SL}(2)\times{\rm GL}(1)_{\rm IIB}
\,,\\
{\bf 27} &\longrightarrow (5,1)_{+4} + (5',2)_{+1} +(10,1)_{-2} + (1,2)_{-5}
\,, 
\label{solvesectionE6B}
\end{split}
\end{equation}
and restricting internal coordinate dependence to the first five coordinates.

The explicit map of the ExFT fields (\ref{ExFTfieldsE6}) into the fields of ten- and eleven-dimensional supergravity
has been worked out in \cite{Hohm:2013vpa,Baguet:2015xha}. Here, we just note that the internal part $g_{mn}$ of the 
higher-dimensional metric can be straightforwardly identified within the components of the matrix ${\cal M}^{MN}$ according to
\begin{equation}
{\cal M}^{MN}\,\partial_M \otimes \partial_N =
({\rm det}\,{g})^{-1/3}\,g^{mn}\,\partial_m \otimes \partial_n 
\,,
\label{MG6}
\end{equation} 
where indices $m, n$ label the derivatives along the physical coordinates 
embedded into the $\partial_M$ according to (\ref{solvesectionE6A}) and (\ref{solvesectionE6B}), respectively.

\subsection{E$_{7(7)}$ ExFT}

The structure of E$_{7(7)}$ exceptional field theory (ExFT) closely parallels the previous construction modulo a few technical 
distinctions. The construction of this theory is based on a split of coordinates into four external 
and 56 internal coordinates, the latter constrained by the section constraint
\begin{equation}
\Omega^{MK}\,(t_\alpha)_K{}^{N}\,\partial_M \Phi_1 \partial_N \Phi_2  = 0 = 
\Omega^{MN}\partial_M \Phi_1 \partial_N \Phi_2
\,,\qquad
\alpha=1, \dots 133
\,,
\label{sectionE7}
\end{equation}
where $\Omega^{MK}$ and $(t_\alpha)_M{}^{N}$ denote the symplectic invariant tensor and
the 133 generators of E$_{7(7)}$, respectively.
The two inequivalent solutions of the section constraint (\ref{sectionE7}) restrict the internal coordinate dependence
of the fields to the six and seven internal coordinates of IIB and $D=11$ supergravity,
respectively.
The bosonic field content of E$_{7(7)}$ ExFT is given by
\begin{equation}
\left\{
g_{\mu\nu}, {\cal M}_{MN}, {\cal A}_{\mu}{}^M, {\cal B}_{\mu\nu\,\alpha},  {\cal B}_{\mu\nu\,M} \right\}
\,,\qquad
\mu, \nu =0, \dots, 3\;,\quad
M=1, \dots, 56\,,
\label{ExFTfieldsE7}
\end{equation}
where the `internal' $56\times56$ metric ${\cal M}_{MN}$ now
parametrizes the coset space ${\rm E}_{7(7)}/{\rm SU}(8)$\,, and can thus also be expressed in terms of a generalised vielbein
\begin{equation} \label{eq:GenMetricE7}
\gM_{MN} = {\cal E}_M{}^{\uM}\, {\cal E}_N{}^{\uN}\, \delta_{\uM\uN} \,,
\end{equation}
where the generalised vielbein, ${\cal E}_M{}^{\uM}$, is now an ${\rm E}_{7(7)}$-valued matrix.
Moreover, apart from two-forms ${\cal B}_{\mu\nu\,\alpha}$ in the adjoint representation of E$_{7(7)}$,
the theory features covariantly constrained two forms ${\cal B}_{\mu\nu\,M}$, subject to algebraic constraints
which parallel the structure of (\ref{sectionE7})
\begin{equation}
0 = \Omega^{MK}\,(t_\alpha)_K{}^{N}\,{\cal B}_{\mu\nu\,M} \, \partial_N \Phi 
= \Omega^{MK}\,(t_\alpha)_K{}^{N}\,{\cal B}_{\mu\nu\,M} {\cal B}_{\rho\sigma\,N} 
\;,\qquad
\alpha=1, \dots 133
\,.
\label{sectionE7B}
\end{equation}

The dynamics of E$_{7(7)}$ ExFT  is most compactly described by a pseudo-Lagrangian
\begin{equation}
 {\cal L}_{\rm ExFT7} = \sqrt{|g|}\, \Big( \widehat{R}
 +\frac{1}{48}\,g^{\mu\nu}{\cal D}_{\mu}{\cal M}^{MN}\,{\cal D}_{\nu}{\cal M}_{MN}
-\frac{1}{8}\,{\cal M}_{MN}{\cal F}^{\mu\nu M}{\cal F}_{\mu\nu}{}^N
 +\sqrt{|g|}{}^{-1}{\cal L}_{\rm top}
-V(g,{\cal M})\Big) \,,
\label{ExFTE7}
\end{equation}
amended by the twisted self-duality equation
\begin{equation}
{\cal F}_{\mu\nu}{}^M =
-\frac12\,\sqrt{|g|}\,\varepsilon_{\mu\nu\rho\sigma}\,\Omega^{MN} {\cal M}_{NK}\,{\cal F}^{\rho\sigma\,K}
\,,
\label{twistedSD}
\end{equation}
for the non-abelian vector field strengths
\begin{equation}
	\begin{split}
{\cal F}_{\mu\nu}{}^M &\equiv
2\, \partial_{[\mu} {\cal A}_{\nu]}{}^M 
-2\,{\cal A}_{[\mu}{}^K \partial_K {\cal A}_{\nu]}{}^M 
-\frac1{2}\left(24\, (t_\alpha)^{MK} (t^\alpha)_{NL}
-\Omega^{MK}\Omega_{NL}\right)\,{\cal A}_{[\mu}{}^N\,\partial_K {\cal A}_{\nu]}{}^L \\
& \quad
 - 12 \,  (t^\alpha)^{MN} \partial_N {\cal B}_{\mu\nu\,\alpha}
-\frac12\,\Omega^{MN} {\cal B}_{\mu\nu\,N}
\,.
\label{F7}
\end{split}
\end{equation}

The various terms in (\ref{ExFTE7}) are defined in complete analogy to \eqref{ExFTE6} above. 
In particular, covariant derivatives are defined as in \eqref{covDE6} where now ${\cal L}_\Lambda$ refers
to generalised internal diffeomorphisms \eqref{gen_diff_M} for the group E$_{7(7)}$ with $\alpha_7=12$, and 
 the projector onto the adjoint representation expressed as
\begin{equation}
\mathbb{P}^K{}_M{}^L{}_N
=
\frac1{24}\,\delta_M^K\delta_N^L
+\frac1{12}\,\delta_M^L\delta_N^K
+(t_\alpha)_{MN} (t^\alpha)^{KL}
-\frac1{24} \,\Omega_{MN} \Omega^{KL}
\,.
\end{equation}
The topological term is defined via
\begin{equation}
d{\cal L}_{\rm top} \propto
24 \,  (t^\alpha)_{M}{}^{N} {\cal F}^M \wedge \partial_N {\cal H}_{\alpha}
+\, {\cal F}^M \wedge {\cal H}_{M}
\,,
\label{dLtopE7}
\end{equation}
in terms of vector and tensor field strengths, while the potential term is still of the universal form \eqref{fullpotential}, now with $\alpha_7=12$\,.
In analogy with \eqref{MG6}, the internal part of the higher-dimensional metric can be identified 
among the components of ${\cal M}^{MN}$ as
\begin{equation}
{\cal M}^{MN}\,\partial_M \otimes \partial_N =
({\rm det}\,{g})^{-1/2}\,g^{mn}\,\partial_m \otimes \partial_n 
\,.
\label{MG7}
\end{equation}

The field equations derived from \eqref{ExFTE6} and \eqref{ExFTE7} reproduce the field equations
of $D=11$ and IIB supergravity, depending on the choice of solution of the section constraint.
Moreover, massive IIA supergravity can be reproduced upon further deformation of the gauge structures 
\cite{Ciceri:2016dmd,Cassani:2016ncu}.

\subsection{Generalised Scherk-Schwarz reduction}
\label{subsec:genSchSch}

One of the powerful applications of the ExFT framework is the description of consistent truncations of higher-dimensional supergravities~\cite{Berman:2012uy,Lee:2014mla,Hohm:2014qga}, i.e.\ truncations to lower-dimensional supergravities such that any solution of the lower-dimensional field equations can be uplifted to a solution of the higher-dimensional field equations. Here, we focus on consistent truncations to maximal supergravities whose field content is precisely of the form \eqref{ExFTfieldsE6} and \eqref{ExFTfieldsE7}, respectively, i.e.\ mirrors the ExFT variables, with fields depending only on the external coordinates.

\subsubsection{Truncation Ansatz}

In terms of the ExFT variables, a consistent truncation to $D=5$ and $D=4$
dimensions, respectively, is described by a reduction Ansatz which on the vector fields takes the form
\begin{equation}
{\cal A}_{\mu}{}^{M}(x,y) ={\cal U}_{\underline{N}}{}^{M}(y)\, A_{\mu}{}^{\underline{N}}(x) \;, 
\label{SchSchA}
\end{equation}
factorizing the dependence on internal and external coordinates into 
an (${\rm E}_{d(d)} \times \mathbb{R}^+$)-valued twist matrix ${\cal U}$ depending on the internal coordinates
and the gauge fields $A_{\mu}{}^{\underline{N}}$ of the lower-dimensional maximal supergravity.
Similarly, external and internal metrics reduce as
\begin{equation}
	\begin{split}
\label{SchSchgM}
g_{\mu\nu}(x,y) &= \rho^{-2}(y)\,g_{\mu\nu}(x)\,,\\
{\cal M}_{MN}(x,y) &= U_{M}{}^{\underline{K}}(y)\,U_{N}{}^{\underline{L}}(y)\,M_{\underline{KL}}(x)\,, 
\end{split}
\end{equation}
respectively, upon decomposing the twist matrix according to
\begin{equation} \label{eq:gU}
{\cal U}_{\underline{M}}{}^N \equiv \rho^{-1} (U^{-1})_{\underline{M}}{}^N
\,,
\end{equation}
into a unimodular matrix $U^{-1} \in {\rm E}_{d(d)}$, and a scale factor $\rho$\,. Finally, the reduction Ansatz for the two-form tensor fields takes the form
\begin{equation}
	\begin{split}
{\rm E}_{6(6)} &:  {\cal B}_{\mu\nu\,M}(x,y) ~=~ 
\,\rho^{-2}(y)\, U_M{}^{\underline{N}}(y)\,B_{\mu\nu\,\underline{N}}(x)
\,,\\[2ex]
{\rm E}_{7(7)} &: 
\left\{
\begin{array}{rcl} {\cal B}_{\mu\nu\,\alpha}(x,y) &=&
	\,\rho^{-2}(y)\, U_\alpha{}^{\underline{\beta}}(y)\,B_{\mu\nu\,\underline{\beta}}(x)
	\;,\\[.5ex]
	{\cal B}_{\mu\nu\,M}(x,y) &=& 
	-2\, \rho^{-2}(y)\,(U^{-1})_{\underline{S}}{}^P(y) \,\partial_M U_P{}^{\underline{R}}(y) (t^{\underline\alpha}){}_{\underline{R}}{}^{\underline{S}}\, B_{\mu\nu\,\underline\alpha}(x) 
	\,,
\end{array}
\right.
\label{SchSchB}
\end{split}
\end{equation}
in ${\rm E}_{6(6)}$ ExFT and  ${\rm E}_{7(7)}$ ExFT, respectively. Here, $U_\alpha{}^{\underline{\beta}}$ denotes the twist
matrix evaluated in the adjoint representation of ${\rm E}_{7(7)}$.
Consistency of the truncation Ansatz (\ref{SchSchA})--(\ref{SchSchB}) is encoded in a set of differential equations 
on the twist matrix which take the universal form
\begin{equation}
\left[\Gamma_{\underline{MN}}{}^{\underline{K}}\right]_{{\cal R}_d} = -\gamma_d\, X_{\underline{MN}}{}^{\underline{K}}
\,,\qquad \Gamma_{\underline{MN}}{}^{\underline{M}} =(1-D)\,\rho^{-1}\partial_{\underline{N}} \rho 
\,,
\label{SchSchConsistency}
\end{equation}
in terms of the algebra valued currents
\begin{equation}
\Gamma_{\underline{M}\underline{N}}{}^{\underline{K}}  \equiv 
(U^{-1})_{\underline{N}}{}^L\,\partial_{\underline{M}} U_L{}^{\underline{K}} \,,\qquad
\partial_{\underline{M}} \equiv {\cal U}_{\underline{M}}{}^N\partial_N
\,.
\label{Gamma}
\end{equation}
Here, $\gamma_d$ are normalization constants given by $\gamma_6=\frac15$, $\gamma_7=\frac17$,
for ${\rm E}_{6(6)}$ ExFT and  ${\rm E}_{7(7)}$ ExFT, respectively. 
$X_{\underline{MN}}{}^{\underline{K}}$ denotes the constant embedding tensor
characterizing the lower-dimensional theory. The projection $[\dots]_{\cal R}$ refers to the projection 
of the rank three tensor $\Gamma_{\underline{MN}}{}^{\underline{K}}$ 
onto the irreducible representation of E$_{d(d)}$ in which the embedding tensor transforms.
For the theories discussed in this paper, these are ${\cal R}_6={\bf 351}$ and ${\cal R}_7={\bf 912}$.

Let us note that, using the explicit form of the projectors in \eqref{SchSchConsistency}
(which can, for example, be found in \cite{deWit:2002vt}), the first of the consistency relations \eqref{SchSchConsistency}
can be explicitly spelled out as
\begin{equation}
-X_{\underline{MN}}{}^{\underline{K}} =
-\alpha_d\,\mathbb{P}_{\underline{L}}{}^{\underline{P}}{}_{\underline{N}}{}^{\underline{K}}\, { \Gamma}_{\underline{PM}}{}^{\underline{L}}
+\frac{\alpha_d}{(D-1)}\,\mathbb{P}_{\underline{M}}{}^{\underline{L}}{}_{\underline{N}}{}^{\underline{K}} \,{ \Gamma}_{\underline{PL}}{}^{\underline{P}}
+{ \Gamma}_{\underline{MN}}{}^{\underline{K}}
\,,
\label{commproj}
\end{equation}
which will be useful in the following. For the E$_{6(6)}$ case, we point out two more useful relations
\begin{equation}
\begin{split}
	5\, { \Gamma}_{\underline{KL}}{}^{[\underline{M}}  d^{\underline{N}]\underline{KL}} 
&=
-X_{\underline{KL}}{}^{\underline{M}} d^{\underline{N}\underline{KL}} 
\,,\\
 { \Gamma}_{\underline{KL}}{}^{(\underline{M}}  d^{\underline{N})\underline{KL}} 
 &=
 -\frac12\, { \Gamma}_{\underline{KL}}{}^{\underline{K}}  d^{\underline{LMN}} 
\,,
\label{dGamma_sym}
\end{split}
\end{equation}
which are obtained from the contraction of \eqref{commproj} with the $d$-tensor and from the E$_{6(6)}$ invariance of the $d$-tensor, respectively.

Every twist matrix solving equations \eqref{SchSchConsistency} defines a consistent truncation
via the reduction Ansatz \eqref{SchSchA}--\eqref{SchSchB}, such that the higher-dimensional field equations factor into
products of twist matrices and the lower-dimensional field equations.
For later use, let us also give the explicit form of the scalar potential induced in the lower-dimensional
gauged supergravities as functions of the embedding tensor 
$X_{\underline{MN}}{}^{\underline{K}}$~\cite{deWit:2004nw,deWit:2007mt}
\begin{equation}
V_{\rm sugra}  =
\frac{1}{2\,\alpha_d}\,{M}^{\underline{MN}}X_{\underline{MP}}{}^{\underline{R}}
\left(X_{\underline{NR}}{}^{\underline{P}} +\gamma_d\,X_{\underline{NT}}{}^{\underline{S}} \,{M}^{\underline{PT}}{M}_{\underline{RS}}
\right)\,.
\label{sugra_potential}
\end{equation}
Let us also recall, that in a given AdS vacuum the relation between AdS length and cosmological constant $\Lambda$ is given by
\begin{equation}
L^2_{\rm AdS}
= -\frac{(D-1)(D-2)}{2\,\Lambda}~=~ -\frac{(D-1)(D-2)}{V_{\rm sugra}}
\,.
\label{AdSL}
\end{equation}

\subsubsection{Generalised Leibniz parallelisability} \label{s:GenPar}
For the purposes of computing the Kaluza-Klein spectrum, it is useful to view the consistent truncation in a slightly different way. In particular, the twist matrix $U \in {\rm E}_{d(d)}$ defines an ($x$-independent) generalised vielbein for the generalised metric, as in equations \eqref{eq:GenMetricE6}, \eqref{eq:GenMetricE7}, i.e.
\begin{equation}
\gM_{MN} = \Delta_{MN} = U_M{}^{\uM}\, U_N{}^{\uN}\, \delta_{\uM\uN} \,, \label{eq:GenVielb}
\end{equation}
and thus fully defines the internal part of the background, i.e.\ the internal metric and fully internal $p$-form field strengths. Moreover, for a consistent truncation, the twist matrix is globally well-defined. Thus, the generalised frame fields $\gU_\uM{}^{M}$, defined using $\rho$ as in \eqref{eq:gU}, defines a collection of nowhere-vanishing generalised vector fields, and the background is called generalised parallelisable, analogous to ordinary parallelise spaces. However, generalised parallelise spaces need not be parallelisable in the ordinary sense, but more generally form coset spaces \cite{Grana:2008yw,Lee:2014mla,Inverso:2017lrz}. 

Finally, it is useful to rephrase the consistency equations \eqref{SchSchConsistency} in terms of the global frame, $\gU_\uM{}{}^M$, as
\begin{equation}
\gL_{{\cal U}_{{\underline{M}} }} {\cal U}_{\underline{N}}  =
X_{\underline{MN}}{}^{\underline{K}} \, {\cal U}_{\underline{K}} 
\,,
\label{UUXU}
\end{equation}
with the action $\gL$ of generalised diffeomorphisms defined by \eqref{covDE6} together with a canonical weight term. Spaces admitting such a generalised frame field are called generalised Leibniz parallelisable spaces and have several important properties. For example, \eqref{UUXU} immediately implies that the vector fields, ${\cal K}_{\uM}$, contained in the generalised frame fields \eqref{eq:gU} according to
\begin{equation}
 \begin{split}
{\cal K}_{\uM}{}^m\,\partial_m &= {\cal U}_{\underline{M}}{}^M\partial_M
\,,
\label{Killing}
 \end{split}
\end{equation}
generate the gauge algebra specified by the embedding tensor, i.e.
\begin{equation} \label{eq:BackAlgebra}
[{\cal K}_{\uM},\, {\cal K}_{\uN}] = X_{\uM\uN}{}^{\uP}\, {\cal K}_{\uP} \,,
\end{equation}
where $[\,,\,]$ denotes the ordinary Lie bracket. Moreover, the vector fields ${\cal K}_{\uM}$ generating the compact part of the gauge group are necessarily Killing vector fields of the background metric that leave the fluxes invariant. This is clear from the expression of the internal Riemannian metric, which can be easily read off from \eqref{eq:GenVielb} and is given by
\begin{equation} \label{eq:BackMetric}
g^{mn} = {\cal K}_{\uM}{}^m\, {\cal K}_{\uN}{}^n\, \delta^{\uM\uN} \,.
\end{equation}

So far we have only discussed the twist matrix $U_M{}^{\uM}$, i.e.\ the background geometry and fluxes around which we define the consistent truncation. However, the consistent truncation Ansatz \eqref{SchSchgM} implies that every space within the truncation is generalised Leibniz parallelisable. To see this, introduce a vielbein for the lower-dimensional gauged supergravity (SUGRA) scalar matrix $M_{\uM\uN}$, i.e.
\begin{equation}
M_{\uM\uN}(x) = {\cal V}_{\uM}{}^{\uA}(x)\, {\cal V}_{\uN}{}^{\uB}(x)\, \delta_{\uA\uB} \,.
\end{equation}
Now we can define a generalised frame field for every internal space obtained by the consistent truncation by dressing the generalised frame field $\gU_{\uM}{}^M$ with the scalar vielbein $\VI_{\uA}{}^\uM$,
\begin{equation} \label{eq:TwistGenFrame}
{\gU}_{\uA}{}^M(x,y) = \VI_{\uA}{}^\uM(x)\, \gU_{\uM}{}^M(y) \,,
\end{equation}
and, equivalently, a generalised vielbein, which entirely encodes the geometry and fluxes,
\begin{equation} \label{eq:TwistedTwist}
{\cal E}_M{}^{\uA}(x,y) = U_M{}^{\uM}(y)\, {\cal V}_{\uM}{}^{\uA}(x) \,,
\end{equation}
such that the generalised metric, \eqref{eq:GenMetricE6} and \eqref{eq:GenMetricE7},
\begin{equation}
\gM_{MN}(x,y) = {\cal E}_M{}^{\uA}(x,y)\, {\cal E}_N{}^{\uB}(x,y)\, \delta_{\uA\uB} = U_M{}^{\uM}(y)\, U_N{}^{\uN}(y)\, M_{\uM\uN}(x) \,,
\end{equation}
takes exactly the form of the truncation Ansatz \eqref{SchSchgM}. Note that the scale factor $\rho$, as in \eqref{eq:gU}, remains unchanged throughout the consistent truncation. Here, and throughout, we will always use the $\uA,\, \uB$ indices to denote objects that are dressed by the scalar vielbein $\VI_{\uA}{}^{\uM}$. 

Since $\VI_\uA{}^\uM$ only depends on the external coordinates $x$, the generalised Lie derivative of the dressed generalised frame fields gives rise to the dressed embedding tensor, often called the $T$-tensor in the gauged SUGRA literature,
\begin{equation} \label{eq:DressedUUXU}
\gL_{\gU_\uA} \gU_\uB = X_{\uA\uB}{}^{\uC}\, \gU_{\uC} \,,
\end{equation}
with
\begin{equation} \label{eq:DressedEmb}
X_{\uA\uB}{}^{\uC} = \VI_\uA{}^\uM\, \VI_\uB{}^\uN\, {\cal V}_\uP{}^\uC\, X_{\uM\uN}{}^{\uP} \,.
\end{equation}
The properties discussed previously now immediately transfer to any background obtain by the consistent truncation. For example, the vector fields making up the dressed generalised frame fields ${\cal K}_{\uA} = \VI_\uA{}^\uM {\cal K}_\uM$ generate the dressed gauge algebra
\begin{equation}
[ {\cal K}_{\uA},\, {\cal K}_{\uB} ] = X_{\uA\uB}{}^{\uC}\, {\cal K}_{\uC} \,.
\end{equation}
In particular, consider some particular vacuum of the lower-dimensional gauged SUGRA theory that we are interested in, specified by the scalar matrix
\begin{equation} \label{eq:ScalarVacuum}
M_{\uM\uN} = \Delta_{\uM\uN} = {\cal V}_{\uM}{}^{\uA}\, {\cal V}_{\uN}{}^{\uB}\, \delta_{\uA\uB} \,.
\end{equation}
The Riemannian metric at this point of the scalar potential can be compactly expressed as
\begin{equation} \label{eq:DefMetric}
g^{mn} = {\cal K}_{\uA}{}^m\, {\cal K}_{\uB}{}^n\, \delta^{\uA\uB} = {\cal K}_{\uM}{}^m\, {\cal K}_{\uN}{}^n\, \Delta^{\uM\uN} \,,
\end{equation}
with $\Delta^{\uM\uN} \Delta_{\uN\uP} = \delta_{\uP}{}^{\uM}$. Equation \eqref{eq:DefMetric} shows how the scalar matrix at the vacuum $M_{\uM\uN} = \Delta_{\uM\uN}$ deforms the internal geometry. Similar expressions can be derived for the fluxes, see for example \cite{Lee:2014mla,Baguet:2015sma}, but are typically lengthier so that we will not give them here.

\section{Fluctuation Ansatz} \label{s:Fluctuation}
We will now show that ExFT leads to a particularly nice description of the linearised fluctuations around a given 10-/11-dimensional background that corresponds to a solution of maximal gauged SUGRA. As we will see in the following, the natural ExFT formulation of these linearised fluctuations leads to a remarkably compact Kaluza-Klein mass matrices for such a background.

\subsection{General linear fluctuations}

We begin by describing general linear fluctuations around a fixed ExFT background with vanishing ${\cal A}_\mu{}^M$, ${\cal B}_{\mu\nu\,M}$. Such a background is just described by a non-trivial generalised metric
\begin{equation}
 \gM_{MN} = \Delta_{MN} \,,
\end{equation}
and an external metric $\mathring{g}_{\mu\nu}$.
The linear fluctuations of the external metric are straightforward and given by
\begin{equation}
g_{\mu\nu} = \rho^{-2} \left( \mathring{g}_{\mu\nu}(x) + h_{\mu\nu}(x,y) \right) \,,  \label{eq:FlucMetric}
\end{equation}
where $\rho^{-2}$ is required to give the ExFT metric, $g_{\mu\nu}$, the right weight, just as in the generalised Scherk-Schwarz Ansatz \eqref{SchSchgM}. For the vector and 2-form fields, $\cA_\mu{}^M$ and $\cB_{\mu\nu\,M}$, we will use the fact that the consistent truncation defines a generalised parallelisation for any background within the truncation via the dressed generalised vielbein $\IU_{\uA}{}^{M}$ in \eqref{eq:TwistedTwist}, as discussed in section \ref{s:GenPar}. In particular, this implies that the matrices $\IU_{\uA}{}^{M}$, seen as a collection of 27 (in the case of $\rm{E}_{6(6)}$) or 56 (in the case of $\rm{E}_{7(7)}$) vector fields, provide a well-defined basis of the generalised tangent bundle. Moreover, the generalised vielbein induces a basis for generalised bundles of any representation of the exceptional group. For example, in the case of $\rm{E}_{6(6)}$, the $U_M{}^{\uA}$ provide a well-defined basis for the $\overline{\mathbf{27}}$-dimensional bundle in which the two-forms $\cB_{\mu\nu,M}$ live. As a result, we can expand any $\cA_\mu{}^M$ and $\cB_{\mu\nu\,M}$ in terms of the basis defined by the background generalised vielbein $U_{\uA}{}^M$, i.e.
\begin{equation}
\begin{split} \label{eq:FlucTen}
\cA_{\mu}{}^{M} &= \rho^{-1}\, \IU_{\uA}{}^M\, \left(A^{KK}\right){\!}_{\mu}{}^{\uA}(x,y) \,, \\
\cB_{\mu\nu\,M} &= \rho^{-2}\, U_{M}{}^{\uA}\, \left(B^{KK}\right){\!}_{\mu\nu\,\uA}(x,y) \,.
\end{split}
\end{equation}

Finally, we turn to the scalar sector, described by the generalised vielbein, ${\cal E}_M{}^{\uA}$, parametrising the coset space ${\rm E}_{d(d)}/H_{d(d)}$. Since ${\cal E}_M{}^{\uA}$ is an ${\rm E}_{d(d)}$ element, a linear fluctuation of the scalar fields is described by an element of the Lie algebra $j_{\uA}{}^{\uB} \in \mathfrak{e}_{d(d)}$, with
\begin{equation}
\delta {\cal E}_M{}^{\uA} = \frac12\,{\cal E}_M{}^{\uB}\, j_{\uB}{}^{\uA}(x,y) \,.
\label{jee}
\end{equation}
However, the fluctuations belonging to $\mathfrak{h}_{d(d)}$ are unphysical, so that we should take $j_{\uA}{}^{\uB} \in \mathfrak{e}_{d(d)} \ominus \mathfrak{h}_{d(d)}$. This implies that
\begin{equation}
j_{\uA\uB} = j_{\uB\uA} \,,
\label{coset}
\end{equation}
where
\begin{equation}
j_{\uA\uB} = j_{\uA}{}^{\uC}\, \delta_{\uB\uC} \,.
\end{equation}
In turn, for the generalised metric \eqref{eq:GenMetricE6}, \eqref{eq:GenMetricE7}, linearised fluctuations are given by
\begin{equation}
\gM_{MN} = {U}_{M}{}^{\uA}\,{U}_{N}{}^{\uB}\, \left( \delta_{\underline{AB}}  + j_{\underline{AB}}(x,y) \right) = 
\Delta_{MN}(y) + {U}_M{}^{\uA}\, {U}_N{}^{\uB}\, j_{\underline{AB}}(x,y) \,. \label{eq:FlucScalar}
\end{equation}

\subsection{Harmonics} \label{s:Harm}

To determine the Kaluza-Klein masses, we now need to expand the fluctuations $h_{\mu\nu}$, $\left(A^{KK}\right){\!}_{\mu}{}^\uA$, $\left(B^{KK}\right){\!}_{\mu\nu\,\uA}$ and $j_{\uA\uB}$ in terms of a complete basis of fields on the internal manifold. One benefit of our approach is already visible. In the ExFT Ansatz, all the linear fluctuations are scalar fields on the internal manifold, such that we only need to find a complete basis of scalar functions, ${\cal Y}^\Sigma$, on the internal manifold. All the tensorial structure of the fluctuations is taken care of by the generalised vielbein, $U_\uA{}^M$, in the fluctuation Ansatz \eqref{eq:FlucTen} and \eqref{eq:FlucScalar}.

We must now choose a good basis of functions ${\cal Y}^\Sigma$ to obtain the Kaluza-Klein spectrum. Since the topology of the compactification is the same for any solution of the lower-dimensional gauged SUGRA, we can choose ${\cal Y}^\Sigma$ to form representations of the largest symmetry group possible, $G_{\rm max}$, which would correspond to the maximally symmetric point of the gauged SUGRA. Note that this maximally symmetric point must not even correspond to a vacuum of the theory, i.e.\ it need not satisfy the equations of motion. Using the ExFT methods, we can choose any internal space corresponding to some configuration of scalar fields of the lower-dimensional supergravity, even if this scalar configuration does not correspond to a minimum of the potential. For example, for the 4-dimensional ${\cal N}=8$, $\SO{8}$ theory, the maximally symmetric point would be the $S^7$ compactification, and we can choose ${\cal Y}^\Sigma$ to form representations of $G_{\rm max} = \SO{8}$ even if we are interested in another solution of the ${\cal N}=8$, $\SO{8}$ theory which breaks the $\SO{8}$ symmetry. As we will show, this choice of ${\cal Y}^\Sigma$ allows us to efficiently compute the Kaluza-Klein spectrum.

The complete basis of functions ${\cal Y}^\Sigma$ must form a representation of the maximal symmetry group. Typically, the consistent truncation is also defined around the maximally symmetric point, such that the generalised frame fields $\gU_{\uM}$, used to construct the consistent truncation \eqref{SchSchgM}, define the maximally symmetric point. Therefore, we have
\begin{equation}
\begin{split}
\gL_{{\cal U}_{\uM}} {\cal Y}^\Sigma &= {\cal U}_{\uM}{}^{M} \partial_M {\cal Y}^\Sigma = {\cal K}_{\uM}{}^m \partial_m {\cal Y}^\Sigma
= - {\cal T}_{\uM}{}^{\Sigma}{}_{\Omega}\, {\cal Y}^\Omega \,,
\end{split}
\label{defT}
\end{equation}
where ${\cal K}_{\uM}$ are the vector fields making up the generalised frame fields, as in section \ref{s:GenPar}. Since the $\gU_\uM$ generate the Lie algebra of $G_{\rm max}$ via \eqref{UUXU}, the matrices ${\cal T}_{\uM}{}^{\Sigma}{}_{\Omega}$, defined by \eqref{defT}, correspond to the generators of $G_{\rm max}$ in the representation of the complete basis of functions ${\cal Y}^\Sigma$. Using the commutator of generalised Lie derivatives, it is straightforward to show that the generators ${\cal T}_{\uM}{}^{\Sigma}{}_{\Omega}$ satisfy the algebra
\begin{equation}
[{\cal T}_{\uM},\, {\cal T}_{\uN}] =  X_{[\uM\uN]}{}^{\uP}\, {\cal T}_{\uP} \,,
\end{equation}
where $X_{\uM\uN}{}^{\uP}$ is the embedding tensor of the lower-dimensional gauged SUGRA, as in \eqref{UUXU}.

In this paper, we will restrict ourselves to theories with compact $G_{\rm max}$\footnote{Note that even if $G_{\rm max}$ is compact, the gauge group of the gauged supergravity may be non-compact. An example of this would be the $D=4$, ${\cal N}=8$ $\mathrm{ISO}(7)$ gauged supergravity, where $G_{\rm max} = \SO{7}$.}, such that the matrices ${\cal T}_{\uM}$ are antisymmetric
\begin{equation}
{\cal T}_{\uM,\Sigma\Omega}=-{\cal T}_{\uM,\Omega\Sigma}
\,,
\end{equation}
and harmonic indices $\Sigma, \Omega$ are raised and lowered with $\delta_{\Sigma\Omega}$\,. 
As we explain in \ref{sec:mass_spin2}, the complete basis of functions ${\cal Y}^\Sigma$ necessarily correspond to the scalar harmonics of the maximally symmetric compactification. Therefore, we will often refer to ${\cal Y}^\Sigma$ as the harmonics.

Now we can exploit the fact that every background obtained by the consistent truncation has a generalised frame field obtained by dressing the maximally symmetric one $\gU_\uM{}^M$ by the scalar matrix~${\cal V}$,
\begin{equation}
 \gU_{\uA}{}^M = \VI_{\uA}{}^{\uM}\, \gU_{\uM}{}^M \,.
\end{equation}
As a result, the generalised vielbein of the background we are interested in has a simple action on the scalar harmonics ${\cal Y}^\Sigma$ of the maximally symmetric point, given by
\begin{equation} \label{eq:DressedT}
 \gL_{{\cal U}_{\uA}} {\cal Y}^\Sigma = - {\cal T}_{\uA}{}^{\Sigma}{}_{\Omega}\, {\cal Y}^\Omega \,,
\end{equation}
where
\begin{equation}
 {\cal T}_{\uA}{}^{\Sigma}{}_{\Omega} = \VI_\uA{}^\uM\, {\cal T}_{\uM}{}^{\Sigma}{}_{\Omega} \,,
\end{equation}
are the generators of $G_{\rm max}$ dressed by the scalar vielbein ${\cal V}$. Their commutator is given by the dressed embedding tensor \eqref{eq:DressedEmb}
\begin{equation}
 [{\cal T}_{\uA},\, {\cal T}_{\uB}] =  X_{[\uA\uB]}{}^{\uC}\, {\cal T}_{\uC} \,.
\end{equation}

For our Kaluza-Klein Ansatz, we now expand the linear fluctuations of the scalar fields, $j_{\uA\uB}$, in terms of the scalar harmonics. This gives
\begin{equation}
\begin{split} \label{eq:KKAnsatz}
\gM_{MN} &= U_{M}{}^{\uA}\,U_{N}{}^{\uB}\, \Big( \delta_{\uA\uB} + \sum_\Sigma \,{\cal Y}^\Sigma\,j_{\uA\uB,\Sigma}(x) \Big) \,, \\
\cA_{\mu}{}^{M} &= \rho^{-1}\, \IU_{\uA}{}^M \sum_\Sigma\, {\cal Y}^\Sigma\, A_{\mu}{}^{\uA,\Sigma}(x) \,, \\
\cB_{\mu\nu\,M} &= \rho^{-2}\, U_{M}{}^{\uA} \sum_\Sigma\, {\cal Y}^\Sigma \, B_{\mu\nu\,\uA,\Sigma}(x) \,, \\
g_{\mu\nu} &= \rho^{-2} \Big( \mathring{g}_{\mu\nu}(x) + \sum_\Sigma\, {\cal Y}^\Sigma h_{\mu\nu,\Sigma}(x) \Big) \,,
\end{split}
\end{equation}
with the sum running over scalar harmonics. From now onwards, we will drop the explicit summation symbol over the scalar harmonics and use the Einstein summation convention instead. As we will see, with this Ansatz for the fluctuations, equations \eqref{eq:DressedT} and \eqref{eq:DressedUUXU} are all the differential information we need to complete determine the Kaluza-Klein spectrum.

\section{Mass matrices}
\label{sec:mass}

In this section, we linearise the field equations in exceptional field theory
with the fluctuation Ansatz (\ref{eq:KKAnsatz}) in order to derive general formulas
for the Kaluza-Klein mass spectrum around the background defined by the generalised metric
\begin{equation}
{\cal M}_{MN}=\Delta_{MN}\quad \Longleftrightarrow\quad M_{\underline{MN}} = \Delta_{\underline{MN}} = {\cal V}_{\underline{M}}{}^{\underline{A}}{\cal V}_{\underline{N}}{}^{\underline{A}} \;.
\end{equation}
As a general rule of notation, when using the flat basis introduced in (\ref{eq:TwistGenFrame}),
we raise, lower, and contract flat indices with $\delta_{\underline{AB}}$\,.

\subsection{Spin-2}
\label{sec:mass_spin2}

Let us start with the spin-2 sector, for which the computation of the Kaluza-Klein spectrum is the most straightforward. The mass spectrum in this sector is also accessible by other universal methods and can be traced back to computing the eigenmodes of a higher-dimensional wave operator depending only on the background geometry~\cite{Constable:1999gb,Brower:1999nj,Csaki:2000fc,Klebanov:2009kp,Bachas:2011xa}.\footnote{
This has e.g.\ been further exploited in~\cite{Richard:2014qsa,Passias:2016fkm,Pang:2017omp,Gutperle:2018wuk,Dimmitt:2019qla}.}
We will explicitly match this result to our approach below.

In the ExFT formulation of supergravity, the mass terms for the spin-2 fluctuations
descend from the universal couplings of the external metric $g_{\mu\nu}$ within ${\cal L}_{\rm pot}$,
\begin{equation}
{\cal L}_{{\rm mass},g} = 
  \frac{1}{4}\,\sqrt{|g|}\left(
  {\cal M}^{MN}\partial_Mg^{\mu\nu}\partial_N g_{\mu\nu}
  +  {\cal M}^{MN}g^{-2}\partial_Mg\,\partial_Ng
  \right)
  \,,
  \label{mass_coup_spin2}
\end{equation}
c.f.\ (\ref{fullpotential}). With the explicit fluctuation Ansatz (\ref{eq:KKAnsatz})
and the action of internal derivatives on the harmonics ${\cal Y}^\Sigma$ expressed in terms of the
${\cal T}_{\underline{M}}$ matrix according to (\ref{defT}) above, the Lagrangian (\ref{mass_coup_spin2})
gives rise to
\begin{equation}
{\cal L}_{{\rm mass},g} \longrightarrow -\frac{1}{4}\,\rho^{2-D}\,{\cal Y}^\Lambda {\cal Y}^\Gamma\,
 {\Delta}^{\underline{MN}}{\cal T}_{\underline{M},\Lambda}{}^\Sigma 
 {\cal T}_{\underline{N},\Gamma}{}^\Omega  
 \,h_{\mu\nu,\Sigma} h^{\mu\nu}{}_{\Omega}
 +\dots \,,
 \label{massspin2A}
\end{equation}
where the ellipses refer to terms carrying traces and divergences of $h_{\mu\nu}$ which play their
role in the explicit realization of the spin-2 Higgs effect
but do not contribute to the final mass matrix.
The latter is read off from \eqref{massspin2A} after comparing the normalization to the linearised Einstein-Hilbert
term from \eqref{ExFTE6}, \eqref{ExFTE7}:
\begin{equation}
\mathbb{M}_{\Sigma\Omega} = 
- {\Delta}^{\underline{MN}}\, \left({\cal T}_{{\underline M}}{\cal T}_{{\underline N}}\right)_{\Sigma\Omega} 
=-  \left({\cal T}_{\underline{A}}{\cal T}_{\underline{A}}\right)_{\Sigma\Omega} 
 \,,
 \label{massSpin2}
\end{equation}
 in the flat basis introduced in \eqref{eq:TwistGenFrame}.

The full system of differential equations for the spin-2 modes also includes the couplings of these modes to
the spin-1 fluctuations via the connection terms in the Einstein-Hilbert term $\widehat{R}$ and to the spin-0
fluctuations via the respective third terms in the ExFT potential (\ref{fullpotential}). 
Upon gauge fixing, they account
for the transfer of degrees of freedom from the massless vector and scalar fluctuations to building the massive spin-2 modes~\cite{Dolan:1983aa,Cho:1992rq,Hohm:2005sc}.
Rather than working out these couplings in detail, the most direct analysis of their contribution proceeds by 
spelling out the relevant gauge symmetries.
Linearising external diffeomorphisms upon expanding their gauge parameter 
in accordance with the fluctuation Ansatz (\ref{eq:KKAnsatz}) as $\xi^\mu=\sum_\Sigma \xi^{\mu\,\Sigma}\,{\cal Y}^{\Sigma}$
induces the action
\begin{equation}
\delta_\xi h_{\mu\nu,\Sigma} = 2\,\partial_{(\mu} \xi_{\nu),\Sigma}\;,\qquad
\delta_\xi A_{\mu}{}^{\underline{M},\Sigma} ={\cal T}_{\underline{M},\Sigma\Omega} \,\xi_{\mu}{}^{\Omega}
\,,
\label{deltaA}
\end{equation}
which can be used as a shift symmetry to explicitly eliminate those vector field fluctuations which couple to 
the spin-2 fluctuations at the quadratic level.
Next, we turn to the scalar fields to identify the corresponding Goldstone modes here. With the gauge transformations 
on the scalar matrix given by (\ref{gen_diff_M}) above,
let us project these transformations onto those which at the linearised level yield shift symmetries to the
scalar fluctuations. I.e.\ we set ${\cal M}_{MN}$ to its background value 
$\Delta_{MN}$ and expand $\Lambda^M$
into harmonics according to the Ansatz (\ref{eq:KKAnsatz}) for the corresponding gauge fields.
After going to the flat basis (\ref{eq:TwistGenFrame}), these transformations can be brought to the form
\begin{equation}
	\begin{split}
\delta  j_{\underline{AB},\Sigma}
 &=
\Big[ \,\Lambda^{\underline{C},\Sigma} \,
 \Gamma_{\underline{CA}}{}^{B}
+\frac{\alpha_d}{D-1}\,\Gamma_{\underline{CD}}{}^{\underline{C}} 
\,\mathbb{P}^{\underline{D}}{}_{\underline{E}}{}^{{\underline{B}}}{}_{{\underline{A}}}\,
\,\Lambda^{\underline{E},\Sigma} 
-\alpha_d\,
\Gamma_{\underline{CD}}{}^{\underline{E}}
\,\mathbb{P}^{\underline{C}}{}_{\underline{E}}{}^{{\underline{B}}}{}_{{\underline{A}}}\,
\,\Lambda^{\underline{D},\Sigma} 
\\
&\quad
\quad+
\frac{\alpha_d}{2}\,{\cal T}_{\underline{C},\Sigma\Omega}
\,\mathbb{P}^{\underline{C}}{}_{\underline{D}}{}^{{\underline{B}}}{}_{{\underline{A}}}\,
\,\Lambda^{\underline{D},\Omega} 
+
\frac{\alpha_d}{2}\,{\cal T}_{\underline{C},\Sigma\Omega}
\,\mathbb{P}^{\underline{C}}{}_{\underline{D}}{}^{{\underline{A}}}{}_{{\underline{B}}}\,
\,\Lambda^{\underline{D},\Omega} 
~~+~ (A \leftrightarrow B)\;
\Big]_{\rm coset}
\\[2ex]
&=
\Big[ 
- 
\Lambda^{\underline{C},\Sigma} 
X_{\underline{C}\underline{A}}{}^{\underline{B}}
-
\Lambda^{\underline{C},\Sigma} 
X_{\underline{C}\underline{B}}{}^{\underline{A}}
+
\alpha_d\,\Lambda^{\underline{A},\Omega} \,{\cal T}_{\underline{B},\Sigma\Omega}
+
\alpha_d\,\Lambda^{\underline{B},\Omega} \,{\cal T}_{\underline{A},\Sigma\Omega}
\;
\Big]_{\rm coset}
\,,
\label{deltaJ}
\end{split}
\end{equation}
where we have used the projector relations (\ref{commproj}) from above.
In addition, the r.h.s.\ of (\ref{deltaJ}) is understood under
projection onto the symmetric coset valued index pairs $(\underline{AB})$, c.f.~(\ref{jee}), (\ref{coset}) above.
These gauge transformations combine the standard Higgs effect (giving mass to the spin-1
vector fields) with the transformations eliminating the Goldstone scalars for the massive spin-2 modes.
To identify the latter, it is sufficient to evaluate (\ref{deltaJ}) for the gauge parameters
corresponding to the vector fields transforming under (\ref{deltaA}). Combining these two formulas,
we find that the scalars affected
by the spin-2 Higgs mechanism are those transforming under
\begin{equation}
\delta j_{\underline{AB},\Sigma}
 =\Pi_{\underline{AB},\Sigma\Omega}\, \Lambda^{\Omega}
\,,
\label{deltaJspin2}
\end{equation}
with a gauge parameter $\Lambda^\Omega$, and the tensor $\Pi$ defined as
\begin{equation}
\Pi_{\underline{AB},\Sigma\Omega}=
\left[
- 
X_{\underline{C}\underline{A}}{}^{\underline{B}}\,
{\cal T}_{\underline{C},\Sigma\Omega} 
+
\alpha_d\,\left({\cal T}_{\underline{A}}
{\cal T}_{\underline{B}}\right)
_{\Sigma\Omega} \;
\right]_{\rm coset}
\,,
\end{equation}
where again the projection on the r.h.s.\ refers to projection of the ${\underline{AB}}$ indices
onto the symmetric coset valued index pairs $(\underline{AB})$.

To sum up, the full system of differential equations for the spin-2 modes also includes their couplings 
to the spin-1 fluctuations singled out by (\ref{deltaA}) and the spin-0 fluctuations defined by (\ref{deltaJspin2}).
Proper gauge fixing will eliminate the lower spin modes in favour of the massive spin-2 excitations. This does not
alter the result (\ref{massSpin2}) for the spin-2 mass matrix, but will have to be taken into account in the
computation of the spin-1 and spin-0 mass spectra, where these Goldstone modes will have to be explicitly eliminated
before calculating the spectrum.

Let us finally compare the mass matrix (\ref{massSpin2}) to the general analysis of \cite{Bachas:2011xa}. There,
it has been shown that upon compactification from ten dimensions around a warped background metric
\begin{equation}
ds^2 = e^{2A(y)}\,\bar{g}_{\mu\nu}(x)\,dx^\mu dx^\nu +
\hat{g}_{mn}(y)\,dy^m dy^n
\,,
\label{metric10}
\end{equation}
the mass spectrum of the spin-2 fluctuations is encoded in the following Laplace equation on the internal space
\begin{equation}
\Box_{{\rm spin 2}}\,\psi ~\equiv~
e^{(2-D)A}\,|\hat{g}|^{-1/2}\,
\partial_m \left(
|\hat{g}|^{1/2}\,\hat{g}^{mn}\,e^{DA}\,\partial_n \psi \right) =
-m^2\,\psi
\,,
\label{spin2-standard}
\end{equation}
where $D$ is the number of external dimensions: $\mu = 0,\dots, D-1$\,.\footnote{
Strictly speaking, reference \cite{Bachas:2011xa} gives the result for $D=4$, but it straightforwardly
generalizes to arbitrary $D$.}
In particular, this spectrum only depends on the internal background geometry. 

Let us compare \eqref{spin2-standard} to the spin-2 mass matrix \eqref{massSpin2} obtained in our framework. For the compactifications described by ExFT, the internal background metric, $\hat{g}^{mn}$, is embedded into the generalised metric, ${\cal M}^{MN}$, according to the universal relation
\begin{equation}
\hat{g}^{mn}\,\partial_m \otimes \partial_n =
|\hat{g}|^{1/(D-2)}\,{\cal M}^{MN}\,\partial_M \otimes \partial_N
\,,
\end{equation}
based on the embedding of the physical coordinates $y^m$ into the ExFT coordinates, c.f.\ \eqref{MG6}, \eqref{MG7}. Similarly, the ExFT embedding of the external metric together with the reduction Ansatz (\ref{SchSchgM})
yields the identification
\begin{equation}
e^{2A(y)}= \rho^{-2}\,|\hat{g}|^{-1/(D-2)}
\,.
\end{equation}
As a result,  the Laplacian on the internal manifold can be rewritten as
\begin{equation}
	\begin{split}
\Box_y \psi &=
|\hat{g}|^{-1/2}
\partial_M
\left(
|\hat{g}|^{D/(2(D-2))}\,{\cal M}^{MN}\,\partial_N \psi
\right) \\
&= e^{(D-2)\,A(y)}
{\cal K}_{\underline{A}}{}^m
\partial_m
\left(  e^{-D A(y)} \,
{\cal K}_{\underline{A}}{}^n\partial_n \psi
\right)
\,,
\end{split}
\end{equation}
where we have used the reduction Ansatz (\ref{SchSchgM}) for the internal metric as well as the identification (\ref{Killing}) of the Killing vector fields within the Scherk-Schwarz twist matrix.

For the operator $\Box_{{\rm spin 2}}$ defined in (\ref{spin2-standard}),
we thus find the explicit action
\begin{equation}
\Box_{{\rm spin 2}} \,\psi =
{\cal K}_{\underline{A}}{}^m
\partial_m
\left( 
{\cal K}_{\underline{A}}{}^n\partial_n \psi
\right)
\,,
\end{equation}
in terms of the Killing vector fields. Combining this with \eqref{defT},
we find the action on the ${\cal Y}^\Sigma$ as
\begin{equation}
\Box_{{\rm spin 2}}\, {\cal Y}^\Sigma =
 \left({\cal T}_{\underline{A}}{\cal T}_{\underline{A}}\right)_{\Sigma\Omega}  \, {\cal Y}^\Omega
~=~
-\mathbb{M}_{\Sigma\Omega}\,{\cal Y}^\Omega
\,,
\end{equation}
showing agreement of the general result \cite{Bachas:2011xa} with the mass matrix \eqref{massSpin2} in ExFT compactifications. Moreover, this also shows that the ${\cal Y}^\Sigma$ are harmonics of the Laplacian \eqref{spin2-standard}, hence our nomenclature for the ${\cal Y}^\Sigma$. In \cite{Dimmitt:2019qla}, a similar form of the spin-2 mass matrix has been proposed for reductions to $D=4$ maximal supergravity.

\subsection{Tensor fields}
\label{subsec:tensors}

Let us move on to the mass spectrum of antisymmetric tensor fields. Their appearance is specific to compactifications
to $D=5$ dimensions, described by E$_{6(6)}$ ExFT. Their field equation in ExFT is obtained from \eqref{ExFTE6}
by variation w.r.t.\ the tensor fields ${\cal B}_{\mu\nu\,M}$ which leads to the first-order
duality equation
 \begin{equation}
d^{PML}\partial_L  \left(e{\cal M}_{MN} {\cal F}^{\mu\nu N}
 +\frac{\sqrt{10}}{6}  \,\varepsilon^{\mu\nu\rho\sigma\tau}\,
  {\cal H}_{\rho\sigma\tau M}\right) = 0
\,,
\label{dualFH}
\end{equation}
with $e=\sqrt{|g|}$\,. To linear order in the fields, the field strength ${\cal F}_{\mu\nu}{}^{M}$ is given by
\begin{equation}
{\cal F}_{\mu\nu}{}^{M} \longrightarrow
2\, {\partial}_{[\mu} {\cal A}_{\nu]}{}^{M}
+10\,d^{MNK} \partial_K  {\cal B}_{\mu\nu\,N}
\,,
\end{equation}
where the last term is responsible for creating the tensor masses in (\ref{dualFH}).
The latter are thus encoded in the differential operator 
$d^{MNK} \partial_K$. Its action on tensor fields obeying the reduction Ansatz (\ref{eq:KKAnsatz}) is computed as
\begin{equation}
	\begin{split}
U_M{}^{\underline{A}}\,d^{MNK} \partial_K {\cal B}_{\mu\nu N} &=
U_M{}^{\underline{A}}\,
d^{MNK} \partial_K \left(\rho^{-2} U_N{}^{\underline{B}} {\cal Y}^\Sigma \right)
{B}^{\mu\nu}{}_{\underline{B},\Sigma} \\
&=
-\frac{\rho^{-1}}{10}\,
\left(
Z^{\underline{AB}}\,{\delta}_{\Omega\Sigma} 
-10\,
d^{\underline{ABC}}      {\cal T}_{\underline{C},\Omega\Sigma} 
\right)
 {\cal Y}^\Omega 
{B}_{\mu\nu\,\underline{B},\Sigma}
\\
&\equiv
\frac{\rho^{-1}}{\sqrt{10}}\,  
 {\cal Y}^\Omega\,
\mathbb{M}^{\underline{A}\Omega,\underline{B}\Sigma}\,{B}_{\mu\nu\,\underline{B},\Sigma}
\,,
\end{split}
\end{equation}
where we have used \eqref{dGamma_sym}, and \eqref{SchSchConsistency}
and moreover defined the constant antisymmetric tensor
\begin{equation}
Z^{\underline{AB}} = 2\,d^{\underline{CDA}}\,
X_{\underline{CD}}{}^{\underline{B}} 
\,,
\label{defZ}
\end{equation}
that encodes the complete information on the embedding tensor in $D=5$ dimensions.

The result of this computation is the antisymmetric mass matrix 
\begin{equation}
\mathbb{M}^{\underline{A}\Sigma,\underline{B}\Omega} =
\frac{1}{\sqrt{10}}\left(
-Z^{\underline{A}\underline{B}}\,\delta_{\Sigma\Omega} + 10\,   d^{\underline{A}\underline{B} \underline{C}}\, {\cal T}_{\underline{C},\Sigma\Omega}
\right)
\,.
\label{massB}
\end{equation}
The first term arises precisely as in the Scherk-Schwarz reduction to $D=5$ dimensions, the second term captures the effect
of internal derivatives acting on the harmonics.
Plugging back (\ref{massB}) into the duality equation (\ref{dualFH}), we find at linear order
\begin{equation}
0 =
d^{PML}\partial_L  \Big(  \rho^{-2} U_M{}^{\underline{A}}
\,{\cal Y}^\Sigma\,
\Big(
2\, {\partial}^{[\mu} A^{\nu]\, \underline{A},\Sigma} 
+\sqrt{10}\, \mathbb{M}^{\underline{A}\Sigma,\underline{B}\Omega}\,
B^{\mu\nu}{}_{\underline{B},\Omega} +\frac{\sqrt{10}}{2}\,  \varepsilon^{\mu\nu\rho\sigma\tau}\,
  {\partial}_{\rho} {B}_{\sigma\tau\, \underline{A},\Sigma}  
  \Big)
  \Big)
\,. 
 \label{dualityFH_lin}
\end{equation}
The ${\partial}^{[\mu} A^{\nu]\, \underline{A},\Sigma} $ terms can be gauge fixed and amount to the spin-1 Goldstone modes
absorbed into the massive tensor fields. 
This results in the five-dimensional first-order equation
\begin{equation}
3\,{\partial}_{[\mu} {B}_{\nu\rho]\, \underline{A},\Sigma} = \frac12\,\varepsilon_{\mu\nu\rho\sigma\tau}\,\mathbb{M}^{\underline{A}\Sigma,\underline{B}\Omega}\, B^{\sigma\tau}{}_{\underline{B},\Omega} \,,
\label{equationMassB}
\end{equation}
describing topologically massive tensor fluctuations with the mass matrix \eqref{massB}.

Let us finally point out that according to \eqref{dualityFH_lin} the entire first-order equation of the tensor fields is yet hit with another mass operator $d^{PML}\partial_L$. Repeating the same calculation for this action shows that the final first-order equation is given by contracting \eqref{equationMassB} with another mass matrix \eqref{massB}. In other words, zero eigenmodes of the mass matrix $M^{\underline{A}\Sigma,\underline{B}\Omega}$ are in fact not part of the physical spectrum as the corresponding modes among the ${ B}_{\mu\nu\,\underline{A},\Sigma}$ are projected out from all field equations.

\subsection{Vector fields}

In E$_{d(d)}$ ExFT, the field equations obtained by varying the Lagrangian
w.r.t.\ the vector fields are of Yang-Mills type (for $d<8$)
\begin{equation}
 {\nabla}_\nu\left( {\cal M}_{MN}\,{\cal F}^{\nu\mu\,N} \right) =
 {\cal I}_{\rm EH}^{\mu}{}_{M}
+ {\cal I}_{\rm sc}^{\mu}{}_{M}
+ {\cal I}_{\rm top}^{\mu}{}_{M}
 \,,
 \label{Aeom}
\end{equation}
where the currents on the r.h.s. denote the contributions from the Einstein-Hilbert term,
the scalar kinetic term, and the topological term, respectively.
As we have discussed in section~\ref{sec:mass_spin2} above, upon linearisation the contributions from 
${\cal I}_{\rm EH}^{\mu}{}_{M}$ only contribute to Higgsing the spin-2 modes and have no impact on the
masses of the physical spin-1 fluctuations. We will deal with these contributions at the very end by projecting the vector
mass matrix on the physical sector invariant under translations \eqref{deltaA}.
The contributions $ {\cal I}_{\rm top}^{\mu}{}_{M}$ from the topological term 
are in general of higher order in the fields and drop out after linearisation. The notable exception is E$_{7(7)}$ ExFT,
where according to \eqref{dLtopE7} this current carries a contribution dual to the field strengths ${\cal H}_\alpha$,
${\cal H}_M$, which by virtue of the derivative of the twisted self-duality equation \eqref{twistedSD} together with 
the Bianchi identity gives rise to a contribution
which equals the l.h.s.\ of \eqref{Aeom} up to sign.

Upon linearisation, we will thus extract the vector mass matrix 
from the r.h.s.\ of the universal equation
\begin{equation}
\Delta_{MN}\, {\nabla}_\nu {\cal F}^{\nu\mu\,N} =
  {\cal J}_{\rm sc}^{\mu}{}_{M}\,\Big|_{\rm lin}
 \,,
 \label{eomvecgen}
\end{equation}
 where the current ${\cal J}_{\rm sc}^{\mu}{}_{M}$ is defined from variation of the scalar kinetic term
\begin{equation}
	\begin{split}
		e\,\delta {\cal A}_{\mu}{}^M  {\cal J}_{\rm sc}^{\mu}{}_{M} &=
		\delta_A \left(
		\frac{1}{4\,\alpha_d}\,e\,{\cal D}^{\mu}{\cal M}^{MN}\,{\cal D}_{\mu}{\cal M}_{MN}
		\right) \\
		&=
		- e\,\delta {\cal A}_{\mu}{}^M  \left(
		\frac1{2\,\alpha_d} \, (J_M)_K{}^L  J^{\mu}{}_{L}{}^K
		+ \,e^{-1}\,\partial_N\left(e J^{\mu}{}_{M}{}^N \right)
		\right)
		\,,
		\label{Jsc}
	\end{split}
\end{equation}
with the currents
\begin{equation}
 (J_{L})_{N}{}^M = {\cal M}_{NK}\, \partial_L {\cal M}^{KM}\;,\quad
J_{\mu\,N}{}^M = {\cal M}_{NK}\, {\cal D}_\mu {\cal M}^{KM}
\,.
\label{currents}
\end{equation}

To linear order in the fluctuations in \eqref{Jsc}, the internal current only contributes its background value, which in the flat basis reads
\begin{equation}
J_{\underline{A},\underline{B}}{}^{\underline{C}} \,\longrightarrow
-\Big(
\Gamma_{\underline{AB}}{}^{\underline{C}}
+\Gamma_{\underline{AC}}{}^{\underline{B}}
\Big)
\,,
\label{Jback}
\end{equation}
with $\Gamma_{\underline{AB}}{}^{\underline{C}}$ from \eqref{Gamma}.
The external current $J_{\mu\,N}{}^M$ carries vector and scalar fluctuations. However, the latter do not contribute to the vector masses but ensure the proper absorption of the scalar Goldstone modes for realizing the spin-1 Higgs mechanism. The vector fluctuations arise from the connection \eqref{covDE6} within $J_{\mu\,N}{}^M$ and split into terms which due to \eqref{UUXU} carry the embedding tensor $X_{\underline{AB}}{}^{\underline{C}}$ together with the contributions from the harmonics following from \eqref{defT}. In the flat basis \eqref{eq:TwistGenFrame}, these take the form 
\begin{equation}
J_{\mu\,\underline{A}}{}^{\underline{B}} \,\Big|_{\rm lin} =
\left(
 - \left(
   X_{\underline C\underline A}{}^{\underline B}
+   
  X_{\underline C\underline B}{}^{\underline A}
  \right)
A_\mu{}^{\underline C,\Sigma}  
+
 \alpha_d
 \left(
 \mathbb{P}_{\underline{A}}{}^{\underline{B}}{}_{\underline{C}}{}^{\underline{D}}
 +\mathbb{P}_{\underline{B}}{}^{\underline{A}}{}_{\underline{C}}{}^{\underline{D}}
\right)
{\cal T}_{\underline{D},\Sigma\Omega} \,  A_\mu{}^{\underline C,\Omega}  
\right)
 {\cal Y}^\Sigma
\,.
\end{equation}
In \eqref{Jsc}, this current also appears under internal derivative according to
\begin{equation}
	\begin{split}
- e^{-1}\,\partial_N\left(e J^{\mu}{}_{M}{}^N \,\Big|_{\rm lin} \right)
&=
\rho^3\,U_M{}^{\underline{A}} \, \mathring{g}^{\mu\nu} \left(
 \frac{1}{D-1}\,
\Gamma_{\underline{BC}}{}^{\underline{B}}  J_{\nu}{}_{\underline{A}}{}^{\underline{C}} 
-\Gamma_{\underline{CA}}{}^{\underline{B}}
 \,J_{\nu}{}_{\underline{B}}{}^{\underline{C}} 
-\partial_{\underline{B}} \left( J_{\nu}{}_{\underline{A}}{}^{\underline{B}} \right)
\right) \Big|_{\rm lin}
\nonumber\\[1ex]
&\stackrel{(\ref{commproj})}{=}
- \frac{\rho^3}{\alpha_d}\, U_M{}^{\underline{A}}\, \mathring{g}^{\mu\nu}
\left(
\left( X_{\underline{AB}}{}^{\underline{C}}
+
\Gamma_{\underline{AB}}{}^{\underline{C}}\right)
J_{\nu}{}_{\underline{C}}{}^{\underline{B}} 
+\alpha_d\,
\partial_{\underline{B}} \left( J_{\nu}{}_{\underline{A}}{}^{\underline{B}} \right)
\right) \Big|_{\rm lin}
\,.
\end{split}
\end{equation}

With these explicit expressions, the two terms on the r.h.s.\ of \eqref{Jsc} combine into
\begin{equation}
{\cal J}_{\rm sc}^{\mu}{}_{M}\,\Big|_{\rm lin} =
-\frac{\rho^3}{\alpha_d}\, U_M{}^{\underline{A}}\, \mathring{g}^{\mu\nu}
\left(
X_{\underline{AB}}{}^{\underline{C}}\,J^\mu{}_{\underline{C}}{}^{\underline{B}}
+
\alpha_d\,\partial_{\underline{B}} \left(J^\mu{}_{\underline{A}}{}^{\underline{B}} \right)
\right)
\Big|_{\rm lin}
\,.
\label{J12}
\end{equation}
Consistently, all $\Gamma_{\underline{AB}}{}^{\underline{C}}$ terms have dropped out and
only the terms carrying the constant embedding tensor $X_{\underline{AB}}{}^{\underline{C}}$
as well as the matrices ${\cal T}_{\underline{A}}$ survive.

Putting everything together, we can write the linearised $D$-dimensional vector field equation \eqref{eomvecgen} as
\begin{equation}
\partial_\nu \partial^{\nu} A^{\mu\,\underline{A},\Sigma}
-\partial_\nu \partial^{\mu} A^{\nu\,\underline{A},\Sigma} 
 =
 \mathbb{M}_{\underline{A}\Sigma,\underline{B}\Omega}\,
  A^{\mu\,\underline{B},\Omega}
 \,,
 \label{eomveclin}
 \end{equation}
where the mass matrix results from collecting all the resulting terms in \eqref{J12} and takes the form
\begin{equation}
	 \begin{split}
\mathbb{M}_{\underline{A}\Sigma,\underline{B}\Omega}&=
\frac1{\alpha_d}\,
\,
X_{\underline{AD}}{}^{\underline{C}}\,
 \left(
   X_{\underline B\underline C}{}^{\underline D}
+   
  X_{\underline B\underline D}{}^{\underline C}
  \right)
  \delta_{\Sigma\Omega}
+
 \left(
   X_{\underline B\underline A}{}^{\underline C}
+   
  X_{\underline B\underline C}{}^{\underline A}
  -X_{\underline{AB}}{}^{\underline{C}}
-X_{\underline{AC}}{}^{\underline{B}}
  \right)
{\cal T}_{\underline{C},\Sigma\Omega} \\
&{} \quad 
-\alpha_d
 \left(
 \mathbb{P}_{\underline{A}}{}^{\underline{C}}{}_{\underline{B}}{}^{\underline{D}}
 +\mathbb{P}_{\underline{C}}{}^{\underline{A}}{}_{\underline{B}}{}^{\underline{D}}
\right)
\left( {\cal T}_{\underline{C}}  {\cal T}_{\underline{D}}\right)_{\Sigma\Omega}
\,.
\label{Mvec_gen}
\end{split}
\end{equation}
The first term of this mass matrix reproduces the known vector mass matrix within $D$-dimensional supergravity.
In particular, it vanishes for compact generators $X_{\underline{KL}}{}^{\underline{N}}$ in accordance with the 
massless vectors from the supergravity multiplet associated with the unbroken gauge symmetries. The result holds for, both, E$_{6(6)}$ ExFT and E$_{7(7)}$ ExFT.

In a final step, we need to project out by hand the spin-1 Goldstone modes absorbed into the massive spin-2 fields.
To this end, we have to project the vector fluctuations to the subsector that remains invariant under the corresponding 
translations \eqref{deltaA}.
In contrast, the spin-1 modes absorbed into massive tensor modes according to the
discussion after \eqref{dualityFH_lin} above, appear as zero eigenvalues of the mass matrix
\eqref{Mvec_gen} and can thus easily be identified.

\subsection{Scalar fields}

It remains to work out the scalar mass spectrum for the fluctuation Ansatz presented above.
Linearising the theories \eqref{ExFTE6}, \eqref{ExFTE7}, the scalar field equations also contain spin-2 and spin-1
contributions implementing the corresponding Higgs mechanisms. These have no impact on the masses of the physical scalars.
We will deal with these contributions at the end by applying an overall projection to the resulting mass matrix.
Ignoring vector and metric fluctuations, the scalar masses are obtained from the linearised field equation
\begin{equation}
	\begin{split}
\Box {\cal M}_{MN}\Big|_{\rm lin}
 &=
 \Big[
-\frac{1}{2}\,J_{M,K}{}^L \, J_{N,L}{}^K
-\alpha_d\,
J_{K,M}{}^L J_{L,N}{}^K  
+{\Delta}^{QL} 
J_{K,Q}{}^K\,J_{L,M}{}^P\,{\Delta}_{PN}\\
& \quad \;\;\;+2\,(D-1) \,{\alpha_d}\,\partial_K\rho\rho^{-1}  \,J_{M,N}{}^K 
-(D-1)\,\partial_K \rho \rho^{-1}\, {\Delta}^{KL}  J_{L,M}{}^P\,{\Delta}_{PN}\\
&{} \quad
\;\;\;+\rho\,
\,{\Delta}^{KL}\,\partial_K \left(\rho^{-1}\,J_{L,M}{}^P\right) {\Delta}_{PN}
-2\, \rho\,{\alpha_d}\,\partial_K\left(\rho^{-1} J_{M,N}{}^K\right) \\
& \quad
\;\;\;  -4\,D\, {\alpha_d}\,\rho^{-2}\,\partial_M\rho\,\partial_N \rho
 +2\,D\,{\alpha_d}\,\rho^{-1}\,\partial_M \partial_N\rho
\; \Big]_{\rm coset, lin}
 \,,
 \label{eom_scalars_lin}
\end{split}
\end{equation}
with the current $J_{M,N}{}^K$ from \eqref{currents}. In addition, the r.h.s.\ is understood as being projected onto
symmetric coset values index pairs $MN$, c.f.~\eqref{jee}, \eqref{coset} above.

The computation is considerably more laborious than the preceding calculations for the tensor and the spin-1 sector.
The latter analyses were facilitated by the manifest covariance of the field equations under generalised diffeomorphisms
which, together with the generalised parallelisability \eqref{UUXU},
allowed for a compact derivation of the corresponding mass matrices in terms of the embedding tensor
and the matrix ${\cal T}_{\underline{A}}$. The scalar field equation, in contrast, is not manifestly invariant 
under generalised diffeomorphisms.
As a consequence, it is lengthier to arrange the numerous contributions resulting from \eqref{eom_scalars_lin}
until the dependence on the internal coordinates factors out.

We may, however, exploit the known structures from gauged supergravity
to reduce the computation to a few relevant terms.
As for the vector mass matrix \eqref{Mvec_gen}, the contributions to the scalar mass matrix from \eqref{eom_scalars_lin}
can be organised into (schematically)
\begin{equation}
\mathbb{M} = XX + X{\cal T} + {\cal T} {\cal T} 
\,,
\label{Mscalar_schematic}
\end{equation}
according to if internal derivatives hit the twist matrices, $U$, or the harmonics, ${\cal T}$, in the fluctuation Ansatz
\eqref{eq:KKAnsatz}.

The $XX$ terms in \eqref{Mscalar_schematic} do not act on the harmonics and by construction coincide with the mass formula
from gauged supergravity for the lowest multiplet. We can thus directly extract these terms from the variation of
the $D$-dimensional supergravity potentials \eqref{sugra_potential} and only focus on the remaining terms.

To this end, we expand the currents $J_{M,N}{}^K$ to linear order in the fluctuations, which in the flat basis \eqref{eq:TwistGenFrame}
takes the form
\begin{equation}
J_{\underline{D},\underline{A}}{}^{\underline{B}} =
-
\,\Big\{
\Gamma_{\underline{DA}}{}^{\underline{B}}
+\Gamma_{\underline{DB}}{}^{\underline{A}}
+ \partial_{\underline{D}} {j}_{\underline{AB}}
+\Gamma_{\underline{DB}}{}^{\underline{E}}\; {j}_{\underline{AE}}
-\Gamma_{\underline{DE}}{}^{\underline{A}}\;{j}_{\underline{EB}}
\Big\}
\,,
 \label{Jlin}
\end{equation}
extending \eqref{Jback}. Next, we expand their derivatives in the flat basis and obtain
\begin{equation}
	\begin{split}
\nabla_{\underline{C}} J_{\underline{D},\underline{A}}{}^{\underline{B}} &\equiv
\rho^{-1} (U^{-1})_{\underline{C}}{}^L (U^{-1})_{\underline{D}}{}^K (U^{-1})_{\underline{A}}{}^M U_N{}^{\underline{B}}\,
\partial_L\left(\rho^{-1}J_{K,M}{}^N\right)\\[1ex]
&=
- \partial_{\underline{C}} \Gamma_{\underline{DA}}{}^{\underline{B}}
- \partial_{\underline{C}} \Gamma_{\underline{DB}}{}^{\underline{A}}
-\Gamma_{\underline{CA}}{}^{\underline{G}}\,\Gamma_{\underline{DG}}{}^{\underline{B}}
-\Gamma_{\underline{CA}}{}^{\underline{G}}\,\Gamma_{\underline{DB}}{}^{\underline{G}}
\\
&{} \quad
-\Gamma_{\underline{CD}}{}^{\underline{G}}\,\Gamma_{\underline{GA}}{}^{\underline{B}}
-\Gamma_{\underline{CD}}{}^{\underline{G}}\,\Gamma_{\underline{GB}}{}^{\underline{A}}
+\Gamma_{\underline{CG}}{}^{\underline{B}}\,\Gamma_{\underline{DA}}{}^{\underline{G}}
+\Gamma_{\underline{CG}}{}^{\underline{B}}\,\Gamma_{\underline{DG}}{}^{\underline{A}}
\\
&{} \quad
-\partial_{\underline{C}}\Gamma_{\underline{DB}}{}^{\underline{F}}\; {j}_{\underline{AF}}
+\partial_{\underline{C}}\Gamma_{\underline{DF}}{}^{\underline{A}}\; {j}_{\underline{FB}}
-\left(
\Gamma_{\underline{CA}}{}^{\underline{F}}\,\Gamma_{\underline{DB}}{}^{\underline{G}}
+\Gamma_{\underline{CG}}{}^{\underline{B}}\,\Gamma_{\underline{DF}}{}^{\underline{A}}
\right)
{j}_{\underline{FG}}\\
&{} \quad
-\left(
\Gamma_{\underline{CD}}{}^{\underline{G}}\,\Gamma_{\underline{GB}}{}^{\underline{F}}
-\Gamma_{\underline{CG}}{}^{\underline{B}}\,\Gamma_{\underline{DG}}{}^{\underline{F}}\right) {j}_{\underline{AF}}
+\left(
\Gamma_{\underline{CD}}{}^{\underline{G}}\,\Gamma_{\underline{GF}}{}^{\underline{A}}
+\Gamma_{\underline{CA}}{}^{\underline{G}}\,\Gamma_{\underline{DF}}{}^{\underline{G}}\right)
{j}_{\underline{FB}}
\\
&{} \quad
-\Gamma_{\underline{DB}}{}^{\underline{F}}\, \partial_{\underline{C}}{j}_{\underline{AF}}
+\Gamma_{\underline{DF}}{}^{\underline{A}}\partial_{\underline{C}} {j}_{\underline{FB}}
- \Gamma_{\underline{CA}}{}^{\underline{G}}\, \partial_{\underline{D}} {j}_{\underline{GB}}
+ \Gamma_{\underline{CG}}{}^{\underline{B}}\,\partial_{\underline{D}} {j}_{\underline{AG}}
\\
&{} \quad
- \Gamma_{\underline{CD}}{}^{\underline{G}}\,\partial_{\underline{G}} {j}_{\underline{AB}}
- \partial_{\underline{C}} \partial_{\underline{D}} {j}_{\underline{AB}}
\,.
\label{derJ}
\end{split}
\end{equation}

Putting everything together, the r.h.s.\ of \eqref{eom_scalars_lin} vanishes on the background
(${j}_{\underline{AB}}\rightarrow0$) as a consequence of the fact that we are linearising the theory
around a stationary point of the gauged supergravity potential. The terms carrying ${j}_{\underline{AB}}$
will precisely recombine into the $XX$ contributions in \eqref{Mscalar_schematic} which we can extract from 
the gauged supergravity describing the lowest multiplet in the absence of higher fluctuations.
The unknown terms in \eqref{Mscalar_schematic} thus exclusively descend from derivative terms,
such that we can restrict the above expansions to
$\partial_{\underline{C}}{j}_{\underline{AB}}$
\begin{equation}
	\begin{split}
J_{\underline{D},\underline{A}}{}^{\underline{B}} &=
-\Gamma_{\underline{DA}}{}^{\underline{B}}
-\Gamma_{\underline{DB}}{}^{\underline{A}}
- \partial_{\underline{D}} {j}_{\underline{AB}}
+\dots \;,\\[1ex]
\nabla_{\underline{C}} J_{\underline{D},\underline{A}}{}^{\underline{B}} &=
-\Gamma_{\underline{DB}}{}^{\underline{F}}\, \partial_{\underline{C}}{j}_{\underline{AF}}
+\Gamma_{\underline{DF}}{}^{\underline{A}}\partial_{\underline{C}} {j}_{\underline{FB}}
- \Gamma_{\underline{CA}}{}^{\underline{G}}\, \partial_{\underline{D}} {j}_{\underline{GB}}
+ \Gamma_{\underline{CG}}{}^{\underline{B}}\,\partial_{\underline{D}} {j}_{\underline{AG}}\\
&{}
- \Gamma_{\underline{CD}}{}^{\underline{G}}\,\partial_{\underline{G}} {j}_{\underline{AB}}
- \partial_{\underline{C}} \partial_{\underline{D}} {j}_{\underline{AB}}
+\dots
\,,
\end{split}
\end{equation}
with the ellipses denoting the terms that do not contribute to the $X{\cal T} + {\cal T} {\cal T}$ terms in 
\eqref{Mscalar_schematic}. Putting this back into \eqref{eom_scalars_lin}, we are left with
\begin{equation}
	\begin{split}
\mathring{\Box} {j}_{\underline{AB}}
 &=
 \Big[
-2\,\Gamma_{\underline{AC}}{}^{\underline{D}} \, \partial_{\underline{B}}{j}_{\underline{DC}}
-2\,\alpha_d\,\Gamma_{\underline{CD}}{}^{\underline{A}} \partial_{\underline{D}}{j}_{\underline{BC}}  
+2\,\Gamma_{\underline{DA}}{}^{\underline{B}}\, \partial_{\underline{C}}{j}_{\underline{DC}}\,
\\
&{} \quad \;
\;\;-2\,\Gamma_{\underline{CB}}{}^{\underline{D}}\, \partial_{\underline{C}}{j}_{\underline{AD}}
+ 2\,\Gamma_{\underline{CD}}{}^{\underline{B}}\,\partial_{\underline{C}} {j}_{\underline{AD}}
\\
&{} \quad \;
\;\;+2\,\alpha_d\,\Gamma_{\underline{AC}}{}^{\underline{D}}\, \partial_{\underline{C}}{j}_{\underline{BD}}
-2\,\alpha_d\,\Gamma_{\underline{AD}}{}^{\underline{B}}\partial_{\underline{C}} {j}_{\underline{DC}}
+2\,\alpha_d\, \Gamma_{\underline{CB}}{}^{\underline{D}}\, \partial_{\underline{A}} {j}_{\underline{DC}}
\\
&{} \quad \;
\;\;+2\,\alpha_d\,\partial_{\underline{C}} \partial_{\underline{A}} {j}_{\underline{BC}}
- \partial_{\underline{C}} \partial_{\underline{C}} {j}_{\underline{AB}}\;\;
\Big]_{\rm coset}
~~+\mbox{``\,$XX\, {j}$ ''} \,.
\end{split}
\end{equation}
Still, the r.h.s.\ is projected onto coset valued index pairs $({\underline{AB}})$\,.

In a final step, we now expand ${j}_{\underline{AB}}$ into harmonics according to \eqref{eq:KKAnsatz}, such that the action
of internal derivatives can be expressed by the ${\cal T}_{\underline{A}}$ matrix.
We also make the coset projection manifest by contracting the entire fluctuation equation
with another coset-valued fluctuation, such that we find the Lagrangian quadratic in fluctuations
\begin{equation}\label{LSF}
	\begin{split}
{\cal L}_{\rm scalar-fluc} &\propto  
{j}_{\underline{AB},\Sigma} \,\mathring{\Box} {j}_{\underline{AB},\Sigma}
- 4\,\Gamma_{\underline{AC}}{}^{\underline{D}} {\cal T}_{\underline{B},\Omega\Sigma}\;
 {j}_{\underline{AB},\Sigma} 
\,{j}_{\underline{DC},\Omega} 
-4\,\alpha_d\,\Gamma_{\underline{AC}}{}^{\underline{B}}{\cal T}_{\underline{D},\Omega\Sigma}\;
{j}_{\underline{AB},\Sigma} \, {j}_{\underline{CD},\Omega} \\
& \quad
-4\,\Gamma_{\underline{CA}}{}^{\underline{B}}{\cal T}_{\underline{C},\Omega\Sigma}\; {j}_{\underline{AD},\Sigma} \,
{j}_{\underline{BD},\Omega} 
-4\,\alpha_d\,\Gamma_{\underline{BC}}{}^{\underline{A}} {\cal T}_{\underline{C},\Omega\Sigma}\; {j}_{\underline{AD},\Sigma} 
\,{j}_{\underline{BD},\Omega} \\
& \quad
- 2\,\alpha_d\, {\cal T}_{\underline{A},\Omega\Lambda}
{\cal T}_{\underline{B},\Lambda\Sigma}\; {j}_{\underline{AD},\Sigma} \, {j}_{\underline{BD},\Omega}
+   {\cal T}_{\underline{C},\Omega\Lambda}
{\cal T}_{\underline{C},\Lambda\Sigma} \;{j}_{\underline{AB},\Sigma}\, {j}_{\underline{AB},\Omega}
~~+\mbox{``\,$XX {j}\,{j}$ ''}
\,.
\end{split}
\end{equation}

The resulting couplings may be further simplified upon repeated use of the projector property \eqref{commproj} together with
\begin{equation}
	\begin{split}
{j}_{\underline{AB},\Sigma} &= \mathbb{P}^{\underline{A}}{}_{\underline{B}}{}^{\underline{C}}{}_{\underline{D}}\,{j}_{\underline{CD},\Sigma}
\,,\\
{\cal T}_{\underline{F},\Sigma\Omega}\;{j}_{\underline{AE},\Sigma}\,{j}_{\underline{BE},\Omega} &=
{\cal T}_{\underline{F},\Sigma\Omega}\;\mathbb{P}^{\underline{A}}{}_{\underline{B}}{}^{\underline{C}}{}_{\underline{D}}\;
{j}_{\underline{DE},\Sigma}\,{j}_{\underline{CE},\Omega}
\,.
\end{split}
\end{equation}
The first of these relations reflects the algebra-valuedness of the fluctuations while the second one
is a consequence of the closure of the commutator on the algebra. As a consequence, all $\Gamma_{\underline{AB}}{}^{\underline{C}}$
in (\ref{LSF}) can be eliminated in favour of the constant embedding tensor $X_{\underline{AB}}{}^{\underline{C}}$, as required for consistency.
Restoring the $XX {j}\,{j}$ terms as obtained from variation of the gauged supergravity potential \eqref{sugra_potential},
the full scalar mass matrix finally reads
\begin{equation}
	\begin{split}
{j}_{\uA\uB,\Sigma}\; \mathbb{M}^{\uA\uB\Sigma,\uC\uD\Omega}\; {j}_{\uC\uD,\Omega} 
 &=
 X_{\uA\uE}{}^{\uF} X_{\uB\uF}{}^{\uE} \;
  {j}_{\uA\uD,\Sigma}\, {j}_{\uB\uD,\Sigma} 
\\
 &\quad + {\gamma_d}
 \left( X_{\uA\uE}{}^{\uF} X_{\uB\uE}{}^{\uF} + X_{\uE\uA}{}^{\uF} X_{\uE\uB}{}^{\uF} + X_{\uE\uF}{}^{\uA} X_{\uE\uF}{}^{\uB}  \right)
  {j}_{\uA\uD,\Sigma}\, {j}_{\uB\uD,\Sigma} 
   \\
  &\quad+  2\,\gamma_d
  \left( X_{\uA\uC}{}^{\uE} X_{\uB\uD}{}^{\uE} - X_{\uA\uE}{}^{\uC} X_{\uB\uE}{}^{\uD} - X_{\uE\uA}{}^{\uC} X_{\uE\uB}{}^{\uD} \right) 
{j}_{\uA\uB,\Sigma}\, {j}_{\uC\uD,\Sigma} 
\\
  & \quad
-4\,
X_{\underline{AC}}{}^{\underline{D}} {\cal T}_{\underline{B},\Omega\Sigma}
\, {j}_{\underline{AB},\Sigma} 
\,{j}_{\underline{CD},\Omega} 
-4\,X_{\underline{CA}}{}^{\underline{B}}{\cal T}_{\underline{C},\Omega\Sigma}
\, {j}_{\underline{AD},\Sigma} \,
{j}_{\underline{BD},\Omega} 
\\
& \quad
+ 2\,\alpha_d\, {\cal T}_{\underline{A},\Omega\Lambda}
{\cal T}_{\underline{B},\Lambda\Sigma}\;
{j}_{\underline{AD},\Sigma} \, {j}_{\underline{BD},\Omega}
 -  {\cal T}_{\underline{C},\Omega\Lambda} {\cal T}_{\underline{C},\Lambda\Sigma} \;
  {j}_{\underline{AB},\Sigma}\, {j}_{\underline{AB},\Omega}
  \,.
\label{scalar_masses}
\end{split}
 \end{equation}

From the mass spectrum obtained by diagonalising this matrix, we still need to project out the Goldstone modes
that render mass to the spin-1 and spin-2 fluctuations, as anticipated at the beginning of this section.
As usual, the Goldstone modes absorbed by  the massive spin-1 fields appear with zero eigenvalue
in \eqref{scalar_masses} and are thus easily identified. The Goldstone modes absorbed into the massive spin-2 fields
in contrast need to be projected out explicitly.
Following the discussion of section~\ref{sec:mass_spin2} above, this can be implemented by projecting
the mass matrix \eqref{scalar_masses} onto those fields that are left invariant under the shift transformations \eqref{deltaJspin2}.

\section{Examples} \label{s:Examples}

We have in the previous section worked out general mass formulas 
\eqref{massSpin2},
\eqref{massB},
\eqref{Mvec_gen},
\eqref{scalar_masses},
for the complete bosonic Kaluza-Klein spectrum around any vacuum lying within a consistent truncation to maximal supergravity. After diagonalising the mass matrices, the corresponding mass eigenstates are identified within ExFT via the fluctuation Ansatz \eqref{eq:KKAnsatz} and can be uplifted to higher dimensions using the dictionary between the ExFT and the original supergravity variables.

In this section, we illustrate these formulas by various examples in four and five dimensions.

\subsection{Vacua of 5-dimensional $\SO{6}$ gauged SUGRA}
In this section, we will apply our mass formulae to the Kaluza-Klein spectra of two vacua of the 5-dimensional $\SO{6}$ gauged supergravity \cite{Gunaydin:1985cu}. This ${\cal N}=8$ supergravity can be obtained by a consistent truncation of IIB supergravity on $S^5$ \cite{Lee:2014mla,Hohm:2014qga,Baguet:2015sma} and contains various interesting vacua, including the ${\cal N}=8$ AdS$_5 \times S^5$ solution of IIB supergravity and the ${\cal N}=2$ $\SU{2} \times \mathrm{U}(1)$-invariant AdS$_5$ vacuum \cite{Pilch:2000ej} dual to the Leigh-Strassler CFT \cite{Leigh:1995ep}. We will use the example of the AdS$_5 \times S^5$ vacuum to demonstrate our formalism, showing that it allows for a compact identification of the BPS multiplets and the IIB fields sourcing the fluctuations. Next, we show, using the example of the ${\cal N}=2$ $\SU{2} \times \mathrm{U}(1)$ vacuum, that our mass formulae also allow to compute the Kaluza-Klein spectrum of vacua for which this was not possible before.

Let us begin by setting up our notation for the consistent truncation of IIB supergravity to the 5-dimensional $\SO{6}$ gauged supergravity. To do this, we use the ${\rm SL}(6)\times {\rm SL}(2)$ basis of ${\rm E}_{6(6)}$ ExFT, in which the fundamental ${\bf 27}$ representation of E$_{6(6)}$ decomposes into
\begin{equation}
 \begin{split}
 {\bf 27} &\longrightarrow (15,1) \oplus (6',2) \,, \\
 \left\{ A^{\underline{M}} \right\} &\longrightarrow \left\{ A^{ab} , A_{a\alpha} \right\}\,,\quad a=1, \dots, 6\,,\quad \alpha=1, 2 \,.
 \label{break27}
 \end{split}
\end{equation}
In this basis, the $d$-symbol takes the form
\begin{equation}
 \begin{split}
 d^{\underline{KMN}} &=
  \left\{
  \begin{array}{l}
	d^{ab}{}_{c\alpha,d\beta} = \frac1{\sqrt{5}}\, \delta^{ab}_{cd}\,\varepsilon_{\alpha\beta}\,,\\
	d^{ab,cd,ef} = \frac1{\sqrt{80}}\,\varepsilon^{abcdef}
	\,.
 \end{array}
 \right.
 \label{S5d}
 \end{split}
\end{equation}

The consistent truncation of IIB supergravity to the ${\rm SO}(6)$ gauged maximal supergravity of \cite{Gunaydin:1985cu} can be described as a generalised Scherk-Schwarz reduction within E$_{6(6)}$ ExFT in the sense discussed in section~\ref{subsec:genSchSch} above, with twist matrices $U_M{}^{\underline{M}}$
constructed from the elementary sphere harmonics on $S^5$
\begin{equation}
{\cal Y}^a  {\cal Y}^a  = 1 \,.
\label{harmonicsY}
\end{equation}
Specifically, the twist matrices are constructed as ${\rm SL}(6)\subset {\rm E}_{6(6)}$ group matrices, given in terms of the round $S^5$ metric $\mathring{g}_{mn} = \partial_m {\cal Y}^a \partial_n{\cal Y}^a$, and the vector field $\mathring{\zeta}^n$ defined by $\mathring\nabla_n\mathring{\zeta}^n = 1$ by
\begin{equation}
 \begin{split}
(U^{-1})_a{}^{\hat{m}} &=
\left\{
(U^{-1})_a{}^{0}  ,\,
(U^{-1})_a{}^{{m}}\right\} \\
&=
\mathring{\omega}^{1/3}
\left\{
\mathring\omega^{-1}\, {\cal Y}^a  ,\,
\mathring{g}^{mn}\partial_n {\cal Y}^a +4\,\mathring{\zeta}^m  {\cal Y}^a
\right\}
\,,
\label{U6}
\end{split}
\end{equation}
where we have introduced the $\SL{6}$ index $\hat{m} = 0, \ldots, 5$. The weight factor is given by $\rho=\mathring{\omega}^{-1/3}$ in terms of the metric determinant $\mathring{\omega}^2 = {\rm det}\, \mathring{g}_{mn}$\,. For computing the Kaluza-Klein masses, we are particularly interested in the vector components, ${\cal K}_{\uM}$, of the generalised parallelisable frame corresponding to the ${\rm E}_{6(6)}$ twist matrices. These are given by \cite{Lee:2014mla,Baguet:2015sma}
\begin{equation} \label{eq:VecS5}
 {\cal K}_{\uM} = \left\{ \begin{array}{l}
 {\cal K}_{ab} = v_{ab} \,, \\
 {\cal K}_{a\alpha} = 0 \,,
 \end{array}\right.
\end{equation}
where
\begin{equation} \label{eq:KillS5}
 v_{ab}{}^m = - \sqrt{2}\, \mathring{g}^{mn} \left( \partial_m {\cal Y}_{[a} \right) {\cal Y}_{b]} \,,
\end{equation}
are the $\SO{6}$ Killing vectors of the round $S^5$.

The resulting  $D=5$ theory is described by an embedding tensor 
\begin{align}
X_{\underline{MN}}{}^{\underline{P}} &=
\left\{
\begin{array}{l}
	X_{ab,cd}{}^{ef} 
	= 2\,\sqrt{2}\,\delta_{[a}^{[e}\delta^{\phantom{e}}_{b][c}\delta_{d]}^{f]}\,, 
	\\[.5ex]
	X_{ab}{}^{c\alpha}{}_{d\beta} =
	-\sqrt{2}\,\delta_{[a}^{c}  \delta^{\phantom{c}}_{b]d}\,\delta_\beta^\alpha  \,,
\end{array}
\right.
\label{S5X}\\[1ex]
\Longleftrightarrow\qquad
Z^{\underline{MN}} &=
\left\{
\begin{array}{l}
	Z^{ab,cd} \,~= 0 \,,
	\\
	Z_{a\alpha,b\beta} ~= \sqrt{10}\,\varepsilon_{\alpha\beta}\,\delta_{ab}
	\,,
\end{array}
\right.
\label{S5Z}
\end{align}
with $Z^{\underline{MN}}$ defined in (\ref{defZ}).

Finally, we need to choose a complete basis of scalar functions in which to expand the ExFT fields via the Kaluza-Klein Ansatz \eqref{eq:KKAnsatz}. As discussed in section \ref{s:Harm}, it is most convenient to choose the complete basis of functions as representations of the maximally symmetric point of the consistent truncation, which in this case corresponds to the round $S^5$. Therefore, we will expand the ExFT fields in terms of the scalar harmonics on the round $S^5$, which are given by polynomials in the elementary $S^5$ harmonics \eqref{harmonicsY} as
\begin{equation}
\begin{split}
\left\{{\cal Y}^\Sigma \right\} = \left\{1,\, {\cal Y}^a,\,   {\cal Y}^{a_1a_2},\,  \ldots,\, {\cal Y}^{a_1\dots a_n},\, \ldots \right\} \,,
\label{harmList}
\end{split}
\end{equation}
where our notation ${\cal Y}^{a_1\dots a_n}\equiv {\cal Y}^{(\!(a_1} \dots {\cal Y}^{a_n)\!)}$ denotes traceless symmetrisation in the elementary harmonics. The index $\Sigma$ thus runs over the tower of symmetric traceless vector representations $[n,0,0]$ of ${\rm SO}(6)$\,. Accordingly, we will refer to the ${\cal Y}^{a_1\dots a_n}$ harmonics as the level $n$ representation.

For the mass formulae, we need to compute the action of the vectors ${\cal K}_{\uM}$, defined by the generalised parallelisation \eqref{eq:VecS5}, on the scalar harmonics ${\cal Y}^\Sigma$. By construction, the $S^5$ Killing vector fields have a linear action on the harmonics, which is block-diagonal level by level and according to \eqref{defT} defines the matrices ${\cal T}_{\underline{M}}$ as the $\SO{6}$ generators in the symmetric $[n,0,0]$ representation. In our conventions,\footnote{
Our summation convention for the harmonic indices $\Sigma, \Omega$ is such that
\begin{equation}
A^{\Sigma}\, B_{\Sigma} = A\,B + A^a\, B_a + A^{a_1a_2}\, B_{a_1a_2}\, + \ldots + A^{a_1\ldots a_n}\, B_{a_1 \ldots a_n} + \ldots
\;.
\end{equation}}
 these take the explicit form
\begin{equation}
 \begin{split}
{\cal T}_{\underline{M},c_1 \ldots c_n}{}^{d_1 \ldots d_n} &= n\, {\cal T}_{\underline{M},(\!(c_1}{}^{(\!(d_1}\, \delta_{c_2}{}^{d_2} \dots \delta_{c_n)\!)}{}^{d_n)\!)} \,,
\label{S5Tn}
\end{split}
\end{equation}
where double parentheses again denote traceless symmetrisation, and
the action on the elementary harmonics is given by
\begin{equation}
{\cal T}_{\underline{M},c}{}^{d} = \left\{ \begin{array}{l}
	{\cal T}_{ab,c}{}^{d} = \sqrt{2}\,\delta_{c[a}\delta_{b]}{}^{d} \,,
	\\[1ex]
	{\cal T}^{a\alpha}{}_{c}{}^{d} = 0 \,.
\end{array}
\right.
\label{S5T}
\end{equation}

We can now straightforwardly apply our mass formulae \eqref{massSpin2}, \eqref{massB}, \eqref{Mvec_gen}, \eqref{scalar_masses} to compute the spectrum of Kaluza-Klein modes around any vacuum of the $\SO{6}$ gauged supergravity. All we have to do is dress the embedding tensor \eqref{S5X}, \eqref{S5Z} and the ${\cal T}$-matrix \eqref{S5T} by the scalar vielbein, ${\cal V}_{\uM}{}^{\uA}$, corresponding to the vacuum we are interested in.

\subsubsection{${\rm AdS}_5, {\cal N}=8$, ${\rm SO}(6)$ vacuum: IIB on $S^5$} \label{s:S5}

In this section, we recompute the Kaluza-Klein spectrum around the maximally supersymmetric AdS$_5\times S^5$ solution of IIB supergravity. This background sits as an ${\cal N}=8$ vacuum within a consistent truncation to the $D=5$ ${\rm SO}(6)$ gauged maximal supergravity of \cite{Gunaydin:1985cu}, which can be described within ExFT, it is thus amenable to our formalism.
Originally, the Kaluza-Klein spectrum on this background has been determined in \cite{Kim:1985ez,Gunaydin:1984fk} by linearising the IIB field equations and exploiting the representation structure of the underlying supergroup SU$(2, 2|4)$, respectively. We will show how to reproduce these results in our formalism.

The AdS$_5\times S^5$ vacuum corresponds to the stationary point at the origin ${M}_{\uM\uN} = \Delta_{\uM\uN} = \delta_{\uM\uN}$ of the scalar potential (\ref{sugra_potential}). Thus, we can choose the scalar vielbein as ${\cal V}_{\uM}{}^{\uA} = \delta_{\uM}{}^{\uA}$. 
We recall, that in the flat basis, indices are raised, lowered, and contracted with $\delta_{\underline{AB}}$ which in the index split
(\ref{break27}) is expressed in terms of $\delta_{ab}$ and $\delta_{\alpha\beta}$, respectively.\footnote{
We denote both, `curved' ${\rm SL}(6)\times{\rm SL}(2)$ indices and
`flat' ${\rm SO}(6)\times{\rm SO}(2)$ indices by $a, b$ and $\alpha, \beta$.
}

The value of the scalar potential at this point is given by
\begin{equation}
V_{\rm sugra}\Big|_0 = -12\qquad
\Longrightarrow\quad L_{\rm AdS} = 1
\,.
\end{equation}

In the original formulation of type IIB supergravity, the computation of the Kaluza-Klein spectrum  around this background requires to expand all fields into the corresponding sphere harmonics. For example, a ten-dimensional scalar field gives rise to a tower of $D=5$ scalar fields
\begin{equation}
\phi(x,y) = \sum_\Sigma\,{\cal Y}^\Sigma(y)\,\varphi_\Sigma(x)
\,,
\end{equation}
according to the tower of scalar harmonics ${\cal Y}^\Sigma$ on the round $S^5$.

On the other hand, in the traditional formulation, ten-dimensional fields with non-trivial transformation under the Lorentz group on $S^5$ in general give rise to several towers of harmonics which are built from products of the elementary harmonics \eqref{harmonicsY} and their derivatives. These can be classified and determined by group theoretical methods~\cite{Salam:1981xd}. E.g.\ the internal part of the ten-dimensional metric gives rise to an expansion
\begin{equation}
g_{mn}(x,y) =
\sum_\Sigma {\cal Y}^\Sigma_{mn}(y)\,g_\Sigma(x)
\,,
\end{equation}
with the harmonics ${\cal Y}^\Sigma_{mn}$ now filling three towers of $\SO{6}$ representations built from the different irreducible components of
\begin{equation}
{\cal Y}_{mn}^{a_1a_2,a_3\dots a_n} \equiv (\partial_m {\cal Y}^{a_1}) (\partial_n {\cal Y}^{a_2}) \, {\cal Y}^{a_3 \dots a_n} \,.
\end{equation}

In our approach, as discussed in section \ref{s:Harm}, we expand all fields in only the scalar harmonics ${\cal Y}^\Sigma$, and the non-trivial Lorentz structure of the Kaluza-Klein fluctuations will arise entirely from multiplying the twist matrices appearing in the fluctuation Ansatz \eqref{eq:KKAnsatz}. We will demonstrate explicitly how this occurs in the following.

\paragraph{Spin-2 fluctuations}

We recall from \eqref{eq:KKAnsatz} that the spin-2 fluctuations directly organise into the scalar harmonics ${\cal Y}^\Sigma$\,. We immediately obtain their mass spectrum from the expression \eqref{massSpin2} for the mass matrix. 
With the ${\cal T}$-matrix given by (\ref{S5Tn})--(\ref{S5T}), this matrix is (up to normalization)
nothing but the quadratic SO(6) Casimir operator, whose eigenvalue on the $[n,0,0]$ symmetric vector representation
is given by
\begin{equation}
\mathbb{M}_{a_1\dots a_n,b_1\dots b_n} =
n(n+4)\,\delta_{(\!(a_1\dots a_n)\!)}{}^{(\!(b_1\dots b_n)\!)}
 \;.
 \label{S5spin2}
 \end{equation}
With the conformal dimension of spin-2 fields given by $\Delta=2+\sqrt{4+m^2L_{\rm AdS}^2}$,
this gives rise to
\begin{equation}
\Delta = 4+n
\,.
 \label{S5spin2D}
\end{equation}

\paragraph{Tensors}
\label{subsec:tensors_S5}

According to the fluctuation Ansatz \eqref{eq:KKAnsatz}, the tensor field fluctuations combine into the tensor product of the fundamental representation \eqref{break27} with the tower of scalar harmonics \eqref{harmList}. We may explicitly spell out the fluctuation coefficients as
\begin{equation}
 \begin{split}
\left\{
B_{\mu\nu\,\underline{A},\Sigma} \right\}
&=
\left\{
B_{\mu\nu\,ab,c_1\dots c_n}, B_{\mu\nu\,a\alpha,c_1\dots c_n}
\right\}
\,.
\label{tensor_comps}
\end{split}
\end{equation}
At level $n$ they fall into ${\rm SO}(6)\times{\rm SO}(2)$ representations 
\begin{equation}
 \begin{split}
B_{\mu\nu\,ab,c_1\dots c_n}~&\in~
[n,1,1]_0\oplus [n,0,0]_0 \oplus [n-1,2,0]_0\oplus [n-1,0,2]_0 \oplus [n-2,1,1]_0
\,, \\
B_{\mu\nu\,a\alpha,c_1\dots c_n}~&\in~
[n+1,0,0]_{\pm\frac12} \oplus [n-1,1,1]_{\pm\frac12} \oplus [n-1,0,0]_{\pm\frac12}
\,,
\label{rep_tensors_S5}
\end{split}
\end{equation}
where we label these representations as $[n_1,n_2,n_3]_j$ by SO(6) Dynkin weights $n_i$ and SO(2) charge $j$.
In terms of the SO(6) vector indices, the different SO(6) representations 
correspond to the symmetrisations
\ytableausetup{smalltableaux}
\bea
{}[n,0,0] &:& {\footnotesize \ydiagram{3} \dots \ydiagram{2}} \;,\nonumber\\[.5ex]
{}[n,1,1] &:& {\footnotesize \ydiagram{3,1} \dots \ydiagram{2}}  \;,\nonumber\\[.5ex]
{}[n,2,0]\oplus [n,0,2] &:& {\footnotesize \ydiagram{3,1,1} \dots \ydiagram{2}} \;.
\eea
Summing over all levels, we thus find for the full spectrum
\begin{equation}
\begin{split}
\left([0,1,1]_0 \oplus[1,0,0]_{\pm\frac12}\right) \otimes
\sum_{n=0}^\infty\, [n,0,0] &=
\sum_{n=0}^\infty\,  \left(2\cdot[n,1,1]_0+[n,0,2]_0+[n,2,0]_0+[n+1,0,0]_0\right)
\\
& \quad
\oplus [0,0,0]_{\pm\frac12} \oplus
\sum_{n=0}^\infty\,  \left( [n,1,1]_{\pm\frac12}+2\cdot [n+1,0,0]_{\pm\frac12}
\right)
\,.
\label{rep_tensors_S5_all}
\end{split}
\end{equation}
Recall, however, from the discussion in section~\ref{subsec:tensors} that within towers, only tensors of non-vanishing mass are part of the physical spectrum.

We may now evaluate the action of the mass matrix 
\eqref{massB} onto the components \eqref{tensor_comps}.
Recall that the tensor mass matrix is antisymmetric and thus has imaginary eigenvalues.
Using the explicit expressions for $d^{\underline{ABC}}$, $Z^{\underline{AB}}$, ${\cal T}_{\underline{A}}$
from (\ref{S5d}), (\ref{S5Z}), and (\ref{S5T}), above, we obtain\footnote{
For the sake of readability, here and in most of the following formulas of this subsection,
we omit the space-time indices $\mu\nu$ which are irrelevant for the diagonalisation problem.}
\begin{equation}
\begin{split}
(\mathbb{M}B)_{ab,c_1\dots c_n}
&=
- \frac12\,n\,  \varepsilon^{abcdef}\, 
B_{cd,e(\!(c_1\dots c_{n-1}} \delta_{c_n)\!)f}
\,,
\\
(\mathbb{M}B)_{a\alpha,c_1\dots c_n}
&=
-(n+1)\,\varepsilon_{\alpha\beta}\,B_{(a|\beta|}{}_{c_1\dots c_n)} 
+  n\,  \varepsilon_{\alpha\beta}\,\,B_{d\beta,d(\!(c_1\dots c_{n-1}} \delta_{c_n)\!)a}
\,.
\label{MB}
\end{split}
\end{equation}
The first equation shows that among the $B_{ab,c_1\dots c_n}$, the only components carrying non-vanishing mass correspond to the $[n,0,2]\oplus[n,2,0]$ representation, antisymmetric in three indices. To compute the corresponding eigenvalue, we explicitly parametrise $B_{ab,c_1\dots c_n}$ as 
\begin{equation}
B_{ab,c_1\dots c_n} =
t^{(\pm)}_{ab(\!(c_1,c_2\dots c_n)\!)}
\,,
\end{equation}
in terms of a tensor $t^{(\pm)}_{abc,d_1\dots d_{n-1}}$, 
(anti-)self-dual in the first three indices
\begin{equation}
t^{(\pm)}_{abc,d_1\dots d_{n-1}} = \pm\frac16\, i\,\varepsilon_{abcdef}\, t^{(\pm)}_{def,d_1\dots d_{n-1}} \,,
\label{sdt3}
\end{equation}
and traceless under any contraction. The action of the mass matrix \eqref{MB} then yields
\begin{equation}
 \begin{split}
(\mathbb{M}B)_{ab,c_1\dots c_n} &= - \frac12\, \varepsilon^{abcde(\!(x_1}\, 
t^{(\pm)}_{cde}{}^{c_2\dots c_n)\!)} - \frac12\, (n-1)\,\varepsilon^{abcde(\!(c_1}\, t^{(\pm)}_{cd}{}^{c_2,c_3\dots c_n)\!)e} \\
&= \pm i\,(n+2)\, t^{(\pm)}_{ab(\!(c_1,c_2\dots c_n)\!)} ~=~\pm i\,(n+2)  \,B_{ab,c_1\dots c_n} \,,
\label{ten_ev1}
\end{split}
\end{equation}
where we have used that $t_{[abc,d]d_2\dots d_{n-1}}=0$, as a consequence of \eqref{sdt3} and tracelessness.

Next, we turn to the second equation of \eqref{MB}. Its r.h.s.\ shows that the action of $\mathbb{M}$ on $B_{a\alpha,c_1\dots c_n}$ is vanishing on the $[n,1,1]$ representation and has eigenvalues $\pm i\,(n+1)$ on the traceless $B_{\mu\nu}{}_{(\!(a}{}^{\alpha}{}_{c_1 \dots c_n)\!)}$. It remains to compute the eigenvalue on the trace part of $B_{a\alpha,c_1\dots c_n}$. To this end, we explicitly parametrize the trace fluctuations as
\begin{equation}
 \begin{split}
B_{a\alpha,c_1\dots c_n} &= \delta_{a(\!(c_1}\,T_{c_2 \dots c_n)\!),\alpha} \\
&= \delta_{a(c_1}\,T_{c_2 \dots c_n),\alpha} - \frac{n-1}{2\,(n+1)}\,\delta_{(c_1c_2} T_{c_3\dots c_n)a,\alpha} \,,
\label{traceB}
\end{split}
\end{equation}
in terms of a tensor $T$, traceless in its SO(6) indices. The latter relates to the trace of $B$ as
\begin{equation}
B_{\alpha,ac_2\dots c_n} =
\frac{(n+2)(n+3)}{n(n+1)}\,
T_{c_2 \dots c_n,\alpha}
\,.
\end{equation}
The action \eqref{MB} then becomes
\begin{equation}
\begin{split}
(\mathbb{M}B)_{a\alpha,c_1\dots c_n} &= -(n+1)\,\varepsilon_{\alpha\beta}\,\delta_{(ac_1}\,T_{c_2 \dots c_n),\beta} +
\frac{n-1}{2}\, \varepsilon_{\alpha\beta}\,\delta_{(c_1c_2} T_{c_3\dots c_na),\beta} \\
& \quad + (n+5)\,\varepsilon_{\alpha\beta}\, \delta_{a(c_1} T_{c_2 \dots c_n),\beta} - \frac{n-1}{n+1}\, \varepsilon_{\alpha\beta}\,  \delta_{a(c_1} T_{c_2\dots c_n),\beta} \\
&= (n+3)\, \varepsilon_{\alpha\beta}\, \Big( \delta_{a(c_1} T_{c_2\dots c_n),\beta} - \frac{(n-1)}{2\,(n+1)}\, \delta_{(c_1c_2} T_{c_3\dots c_n)a,\beta} \Big) \,,
\end{split}
\end{equation}
with eigenvalue $\pm i\,(n+3)$\,.

\begin{table}[bt]
\centering
\begin{tabular}{c|c|c|c}
fluctuation & representation & $m^2L_{\rm AdS}^2$ & $\Delta$
\nonumber\\
\hline
$B_{\mu\nu}{}_{ab,c_1c_2\dots c_n}$ &
$[n-1,0,2]_0\oplus[n-1,2,0]_0$  
& $(n+2)^2$ & $n+4$
\nonumber\\
$B_{\mu\nu\,(\!(a}{}^{\alpha}{}_{c_1\dots c_n)\!)}$ &
$[n+1,0,0]_{\pm\frac12}$
& $(n+1)^2$ & $n+3$
\nonumber\\
$B_{\mu\nu}{}_{ b\alpha ,bc_2\dots c_n}$ &
$[n-1,0,0]_{\pm\frac12}$
& $(n+3)^2$ & $n+5$
\nonumber\\
\end{tabular}
\caption{Masses of tensor fluctuations at level $n$\,.
The conformal dimension is given by $\Delta=2+|m|L_{\rm AdS}$\,.}
\label{tab:masses_tensors}
\end{table}

We summarize the result for all non-vanishing tensor masses at level $n$ in Table~\ref{tab:masses_tensors}. It shows that at level $n$ the tensor spectrum contains three different representations which all come with different masses. In particular, the representations $[n+1,0,0]_{\pm\frac12}$ and $[n-1,0,0]_{\pm\frac12}$ directly correspond to mass eigenstates. In contrast, computing the Kaluza-Klein spectrum in terms of the original IIB variables requires diagonalisation of a coupled system of equations mixing components of different higher-dimensional fields~\cite{Kim:1985ez}. %
The fluctuation Ansatz~(\ref{eq:KKAnsatz}) precisely solves this diagonalisation problem: the mass eigenstates organise according to the scalar tower of harmonics and mix into the IIB fields upon multiplication with the twist matrix $U_M{}^{\underline{A}}$. Let us make this explicit. 
Table~\ref{tab:masses_tensors} shows that the same representation
$[k,0,0]_{\pm\frac12}$ appears twice within the massive tensor fluctuations
as
\begin{equation}
 \begin{split}
b^+_{c_1\dots c_k,\alpha} &\equiv
B_{b\alpha,bc_1\dots c_k}\,,\\
b^-_{c_1\dots c_k,\alpha}&\equiv
 B_{(\!(c_1}{}^{\alpha}{}_{c_2\dots c_k)\!)}
= 
B_{(c_1}{}^{\alpha}{}_{c_2\dots c_k)}
-\frac{k-1}{2\,(k+1)}\, B_{d}{}^{\alpha}{}_{d(c_1 \dots c_{k-2}} \delta_{c_{k-1}c_k)}
\,,
\label{bpm_ten}
\end{split}
\end{equation}
at levels $n=k+1$, and $n=k-1$, respectively, for which we read off the mass eigenvalues $(k+4)^2$, and $k^2$,
respectively. This precisely reproduces the result of~\cite{Kim:1985ez}.

To identify the higher-dimensional origin of the mass eigenstates, we need to combine this result with the dictionary between the ExFT fields and the fields of IIB supergravity \cite{Baguet:2015xha,Baguet:2015sma}. For the original IIB 2-form $C_{\mu\nu}{}^\alpha$ and in combination with the fluctuation Ansatz~(\ref{eq:KKAnsatz}), this gives rise to an expansion  
\begin{equation}
C_{\mu\nu}{}^\alpha =
{\cal Y}^{a}
\sum_{n=0}^\infty
{\cal Y}^{c_1 \dots c_n}\,
B_{\mu\nu}{}^{a\alpha,c_1 \dots c_n}(x)
\,,
\label{towerC}
\end{equation}
where the ${\cal Y}^a$ prefactor descends from the twist matrix $U_M{}^{\underline{A}}$, and the terms under the sum correspond to the scalar tower of harmonics which fall into mass eigenstates.
Expanding the product of harmonics in \eqref{towerC} according to
\begin{equation}
{\cal Y}^{a}{\cal Y}^{c_1 \dots c_n}
=
{\cal Y}^{ac_1 \dots c_n} + \frac{n}{2\,(n+2)} \, \delta^{a(\!(c_1} {\cal Y}^{c_2 \dots c_n)\!)} \,,
\end{equation}
we find for the expansion of the IIB 2-form
\begin{equation}
\begin{split}
C_{\mu\nu}{}^\alpha &= \sum_{n=0}^\infty \Big( {\cal Y}^{ac_1 \dots c_n} \,b^-_{\mu\nu}{}^{ac_1 \dots c_n,\alpha} + \frac{n}{2\,(n+2)} \,{\cal Y}^{c_2 \dots c_n} \, b^+_{\mu\nu}{}^{c_2 \dots c_n,\alpha} \Big) \\
&= \frac16 \, b^+_{\mu\nu}{}^{\alpha} ~+~ \sum_{k=1}^\infty {\cal Y}^{c_1 \dots c_k} \left( b^-_{\mu\nu}{}^{c_1 \dots c_k,\alpha} + \frac{k+1}{2\,(k+3)} \, b^+_{\mu\nu}{}^{c_1 \dots c_k,\alpha} \right) \,,
\label{expB1}
\end{split}
\end{equation}
mixing in its fluctuations different mass eigenstates. A similar computation for the components of the IIB 6-form ${C}_{\mu\nu\, klmn}{}^{\alpha}$, gives rise to its expansion into different linear combinations of the same objects \eqref{bpm_ten} according to
\begin{equation}
 \begin{split}
{C}_{\mu\nu\, lmnp}{}^{\alpha}&=
\sum_{k=1}^\infty
\mathring\omega_{lmnpq}\,
\partial^q {\cal Y}^{c_1} {\cal Y}^{c_2 \dots c_k} \left(
b^-_{\mu\nu}{}^{c_1c_2 \dots c_k,\alpha}
-\frac{k\,(k+1)}{2\,(k+2)^2\,(k+3)}\, 
b^+_{\mu\nu}{}^{c_1 \dots c_k,\alpha}
\right)
\\
&\quad
+4\,\mathring\omega_{lmnpq}\,
\mathring\zeta^{q} 
\,C_{\mu\nu}{}^{\alpha}
\,,
\label{expB2}
\end{split}
\end{equation}
where again the mixing of different mass eigenstates originates from multiplying out
the harmonics from the twist matrix and the scalar tower of harmonics.

\paragraph{Vectors}

We now perform the corresponding computation for the vector spectrum by evaluating the mass 
matrix \eqref{Mvec_gen}.
According to the fluctuation Ansatz~ \eqref{eq:KKAnsatz}, the vector fluctuations organise into
the same ${\rm SO}(6)\times{\rm SO}(2)$ representations as the tensor
fluctuations, which we explicitly denote as
\begin{equation}
 \begin{split}
A_{\mu}{}^{ab,c_1\dots c_n}~&\in~ 
[n,1,1]_0\oplus [n,0,0]_0 \oplus [n-1,2,0]_0\oplus [n-1,0,2]_0 \oplus [n-2,1,1]_0
\,,
\nonumber\\
A_{\mu}{}^{a\alpha,c_1\dots c_n}~&\in~
[n+1,0,0]_{\pm\frac12} \oplus [n-1,1,1]_{\pm\frac12} \oplus [n-1,0,0]_{\pm\frac12}
\,.
\label{vector_comps}
\end{split}
\end{equation}
For the $S^5$ background, the general vector mass matrix  \eqref{Mvec_gen} simplifies drastically since the generators $X_{\underline{A}}$ are compact: $X_{\underline{AB}}{}^{\underline{C}}=-X_{\underline{AC}}{}^{\underline{B}}$. As a consequence, the action of the mass matrix on the vector fluctuations reduces to
\begin{equation}
(\mathbb{M}A)^{\underline{A}\Sigma}=
-6 \,
 \left(
 \mathbb{P}_{\underline{A}}{}^{\underline{C}}{}_{\underline{B}}{}^{\underline{D}}
 +\mathbb{P}_{\underline{C}}{}^{\underline{A}}{}_{\underline{B}}{}^{\underline{D}}
\right)
{\cal T}_{\underline{D},\Lambda\Omega} {\cal T}_{\underline{C},\Sigma\Lambda} \,  
\,  A^{\underline{B}\Omega}
+
\frac{8}{3}\,
{\cal T}_{\underline{A},\Sigma\Lambda}
{\cal T}_{\underline{B},\Lambda\Omega}
\,  A^{\underline{B}\Omega}
\,.
\label{MAA0}
\end{equation}
Evaluating the r.h.s., we find for the adjoint projector \eqref{PadjE6} with \eqref{S5d}
\begin{equation}
 \begin{split}
\mathbb{P}_{a\alpha}{}^{cd}{}_{b\beta}{}^{ef}+
 \mathbb{P}_{cd}{}^{a\alpha}{}_{b\beta}{}^{ef}
  &=
 \frac16\left(\delta_{ab} \delta_{c[e}\delta_{f]d}
+
2\,\delta_{b[c}\delta_{d][e}\delta_{f]a}\right)
\delta_{\alpha\beta}
-\frac1{12}\,\varepsilon_{abcdef} \,\varepsilon_{\alpha\beta}
\,,\\[2ex]
 \mathbb{P}_{ab}{}^{ef}{}_{cd}{}^{gh}
 +\mathbb{P}_{ef}{}^{ab}{}_{cd}{}^{gh}
 &=
\frac1{9}\,\delta_{ab}^{ef}\delta_{cd}^{gh}
 + \frac16\,\delta_{gh}^{ab}\delta_{ef}^{cd} + \frac16\,\delta_{gh}^{ef}\delta_{ab}^{cd}
-\delta_{abcd}^{efgh}
-\delta_{efcd}^{abgh}\;.
\end{split}
\end{equation}
Moreover, the product of ${\cal T}$ matrices takes the explicit form
\begin{equation}
 \begin{split}
 {\cal T}_{cd,c_1\dots c_n\Lambda} \,  {\cal T}_{ef,\Lambda\Omega}
\, A^{\Omega}
&=
-2\,n\,(n-1)\,
\delta_{[c}{}^{(\!(c_1} \, \,A^{c_2\dots c_{n-1}}{}_{d] [e} \delta_{f]}{}^{c_n)\!)} \\
&\quad + n  \,A^{f(\!(c_1\dots c_{n-1}}\,\delta^{c_n)\!)e}{}_{cd} - n  \,A^{e(\!(c_1\dots c_{n-1}}\,\delta^{c_n)\!)f}{}_{cd}
\,.
\end{split}
\end{equation}
Evaluating \eqref{MAA0} on the components \eqref{vector_comps}, we then find after some computation
\begin{equation}
\begin{split}
(\mathbb{M}A)^{ab,c_1\dots c_n}&=
2\,n\,
A^{ab,c_1,c_2\dots c_n}
+2\,n^2\,
A_{[a}{}^{(\!(c_1,c_2\dots c_n)\!)}{}_{b]}
\\
&\quad
+4\,n\,(n-1)\,
A^{d(\!(c_1,c_2\dots c_{n-1}}{}_{d [a}\,\delta_{b]}{}^{c_n)\!)}
-2\,n^2\,
\delta_{[a}{}^{(\!(c_1}  \,A_{b]d}{}^{c_2\dots c_n)\!)d}
\,,
\\[1ex]
(\mathbb{M}A)^{a\alpha,c_1\dots c_n}&=
 n\,(n+3)\,
 A^{a\alpha,c_1\dots c_n }
- n\,(n+3)\,
 A^{(\!(c_1|\alpha|,c_2\dots c_n)\!) a }
 \\
 &\quad
    -n\, (n-1)\,
 A^{b\alpha,b(\!(c_1\dots c_{n-1}} \delta^{c_n)\!)a}
\,.
\label{MA2}
\end{split}
\end{equation}
The second equation shows that for the traceless part in $A^{a\alpha,c_1\dots c_n}$ only the $[n-1,1,1]$ contribution carries a mass whereas the fully symmetric part in the $[n+1,0,0]$ remains massless. Indeed, the latter states are absorbed as Goldstone modes into the corresponding massive tensor excitations, c.f.~Table~\ref{tab:masses_tensors}. To determine the mass eigenvalue of the $[n-1,1,1]$ vectors, we evaluate the second equation of \eqref{MA2} for components satisfying $A^{(a|\alpha|,c_1\dots c_n)}=0$ together with tracelessness and obtain
\begin{equation}
(\mathbb{M}A)^{a\alpha,c_1\dots c_n}=
(n+1)\,(n+3)\,
 A^{a\alpha,c_1c_2\dots c_n}
\,.
\end{equation}

It remains to compute the masses of the trace modes $A^{b\alpha,bc_2\dots c_n}$. Since these states serve as Goldstone modes for the corresponding $[n-1,0,0]$ massive tensors of ~Table~\ref{tab:masses_tensors}, they must appear massless. As a consistency check of our formulas, this can indeed explicitly be verified upon parametrizing the fluctuations as 
\begin{equation}
A^{a\alpha,c_1\dots c_n} =
\delta^{a(\!(c_1}\,T^{c_2 \dots c_n)\!),\alpha}
= \delta^{a(c_1}\,T^{c_2 \dots c_n),\alpha}
- \frac{n-1}{2\,(n+1)}\,\delta^{(c_1c_2} T^{c_3\dots c_n)a,\alpha} \,,
\end{equation}
with a traceless tensor $T$, just as \eqref{traceB} above, and evaluating the action \eqref{MA2}.

We now turn to the first equation of \eqref{MA2}. Its first line shows that within the traceless part of $A^{ab,c_1,c_2\dots c_n}$, the $[n-1,2,0]\oplus[n-1,0,2]$ representations remain massless as required by consistency (they are the Goldstone modes for the corresponding massive tensors, c.f.~Table~\ref{tab:masses_tensors}). In turn, we can compute the mass of the remaining $[n,1,1]$ representation by parametrising the corresponding fluctuations as
\begin{equation}
A^{ab,c_1,c_2\dots c_n} = t^{[a,b]c_1\dots c_n} \,,
\end{equation}
with a tensor $t$, traceless, and symmetric in its last $n+1$ indices. The action \eqref{MA2} then takes the form
\begin{equation}
 \begin{split}
(\mathbb{M}A)^{ab,c_1\dots c_n} &= 2\,n A^{ab,c_1c_2\dots c_n} + 2\,n^2 A_{[a}{}^{(\!(c_1,c_2\dots c_n)\!)}{}_{b]} \\
&= n\,(n+2)\,t^{[a,b]c_1\dots c_n}~=~ n\,(n+2)\,A^{ab,c_1\dots c_n} \,.
\end{split}
\end{equation}
Finally, we compute the action \eqref{MA2} on the trace parts of $A^\mu{}^{ab,c_1\dots Cc_n}$ by parametrising these as
\begin{equation}
 \begin{split}
A^\mu{}^{ab,c_1\dots c_n} 
&=
\delta^{a(\!(c_1}\, T^{c_2\dots c_n)\!),b}
-\delta^{b(\!(c_1}\, T^{c_2\dots c_n)\!),a}
\,,
\\
\Rightarrow
A^\mu{}^{ab,ac_2\dots c_n} 
&=
\frac{5+4n+n^2}{n(n+1)}\, T^{c_2 \dots c_n ,b}
+\frac{(n-1)}{n(n+1)}\ T^{b (c_2 \dots c_{n-1},c_n)}
\,,
\label{AinTT}
\end{split}
\end{equation}
in terms of a trace-free tensor $T^{c_2\dots c_n,a}$, symmetric in its first $n-1$ indices.

For the $[n-2,1,1]$ representation, we further impose that $T^{(C_2 \dots C_n , A)}=0$ and explicit evaluation of \eqref{MA2} after some computation turns into 
\begin{equation}
(\mathbb{M}A)^{ab,c_1\dots c_n}=
 (2+n)\,(4+n)\, A^{ab,c_1\dots c_n}
\;.
\end{equation}
The $[n,0,0]$ representation is described by \eqref{AinTT} with a fully symmetric $T^{c_2\dots c_n,a}$, however the mass for this representation is irrelevant as the corresponding modes are the ones absorbed into the massive spin-2 excitations. As discussed in section~\ref{sec:mass_spin2}, they have to projected out from the physical spectrum.

\begin{table}[bt]
\centering
\begin{tabular}{c|c|c|c}
fluctuation & representation & $m^2L_{\rm AdS}^2$ & $\Delta$
\nonumber\\
\hline
$A_\mu{}^{a(\!(b,c_1c_2\dots c_n)\!)}$ &
$[n,1,1]_0$ & 
$n\,(n+2)$ & $n+3$
\nonumber\\
$A_\mu{}^{ab,c_2c_3\dots c_n b}$ &
$[n-2,1,1]_0$ & 
$(n+2)(n+4)$ & $n+5$
\nonumber\\
$A_\mu{}^{\alpha a,c_1\dots c_n}$ &
$[n-1,1,1]_{\pm\frac12}$ 
& $(n+1)(n+3)$ & $n+4$
\nonumber\\
\end{tabular}
\caption{Masses of vector fluctuations at level $n$\,.
The conformal dimension is given by $\Delta=2+\sqrt{1+m^2L_{\rm AdS}^2}$\,.}
\label{tab:masses_vectors}
\end{table}

We summarize the result for the massive vector fluctuations in Table~\ref{tab:masses_vectors}. At level $n$ the vector spectrum contains three different mass eigenstates in different representations. Summing over all levels, the representation $[k-1,1,1]_0$ appears twice within the massive vector fluctuations as
\begin{equation}
 \begin{split}
a_+^{a,c_1\dots c_k} &\equiv A^{ab,c_1\dots c_k b}\big|_{[k-1,1,1]}\,, \\
a_-^{a,c_1\dots c_k} &\equiv A^{a(\!(c_1,c_2\dots c_k)\!)}\big|_{[k-1,1,1]} \,,
\label{apm_vec}
\end{split}
\end{equation}
at levels $n=k+1$, and $n=k-1$, respectively, for which we read off the mass eigenvalues $(k+3)(k+5)$, and $k^2-1$, respectively. This precisely reproduces the result of~\cite{Kim:1985ez} (c.f.\ their equation (2.27)).

To identify the higher-dimensional origin of these mass eigenstates, we again appeal to the dictionary between the ExFT fields and the fields of IIB supergravity \cite{Baguet:2015xha,Baguet:2015sma}. The vector fluctuations $A^{ab,c_1\dots c_n}$ descend from the off-diagonal part $A_\mu{}^m$ of the 10D metric and components of the 4-form as
\begin{equation}
 \begin{split}
{A}_\mu{}^{m}(x,y) &= \sqrt{2}\,\partial^m{\cal Y}^a \, {\cal Y}^{b} \sum_{n=0}^\infty {\cal Y}^{c_1 \dots c_n}\, A_\mu^{ab,c_1 \dots c_n}(x) \,,\\
{A}_{\mu\,klm}(x,y) &=  \frac12\, \mathring\omega_{klmpq} \, \partial^p {\cal Y}^a \,\partial^q {\cal Y}^b \sum_{n=0}^\infty {\cal Y}^{c_1 \dots c_n}\, A_\mu^{ab,c_1 \dots c_n}(x)
\,.
\end{split}
\end{equation}
Again, the sum corresponds to the tower of scalar harmonics while the prefactor
comes from the twist matrix $(U^{-1})_{\underline{A}}{}^M$ in \eqref{eq:KKAnsatz}. A computation analogous to the one for the tensor fields in section~\ref{subsec:tensors_S5}, expanding the products of harmonics and rearranging the terms in the tower, yields
\begin{equation}
 \begin{split}
{A}_\mu{}^{m}(x,y) &= \sqrt{2}\,\sum_{n=0}^\infty \left( \partial^m{\cal Y}^a  {\cal Y}^{bc_1 \dots c_n} \, a_-^{a,bc_1\dots c_{n}} + \frac{n}{2(n+2)}\,  \partial^m{\cal Y}^a  {\cal Y}^{c_2 \dots c_n} a_+^{a,c_2\dots c_{n}}  \right)  \,, \\
&= \frac{\sqrt{2}}{6}\,  \partial^m{\cal Y}^a \,a_+^{a} + \sqrt{2}\,\sum_{k=1}^\infty \partial^m{\cal Y}^a  {\cal Y}^{c_1 \dots c_k} \left( a_-^{a,c_1\dots c_{k}} + \frac{k+1}{2\,(k+3)}\, a_+^{a,c_1\dots c_{k}} \right) \,,
\label{expA1}
\end{split}
\end{equation}
and
\begin{equation}
\begin{split}
{A}_{\mu\,mpq}(x,y) &=
\mathring\omega_{mpqrs}\, \sum_{n=0}^\infty \Big(\, \frac{n+1}{n+2} \, \partial^r {\cal Y}^a \partial^s {\cal Y}^b {\cal Y}^{c_1 \dots c_n}\, a_-^{a,bc_1\dots c_{n}} \\
& \quad\qquad\qquad\qquad - \frac{n(n-1)}{2(n+2)^2}\, \, \partial^r {\cal Y}^a 
\partial^s {\cal Y}^{c_2} {\cal Y}^{c_3 \dots c_n}\, a_+^{a,c_2\dots c_{n}} \Big) \\
&= \mathring\omega_{mpqrs}\, \sum_{k=1}^\infty \partial^r {\cal Y}^a \partial^s {\cal Y}^{c_1 \dots c_k}\, \left( \frac{1}{k+1} \, a_-^{a,c_1\dots c_{k}} -  \frac{k+1}{2(k+3)^2}\, \,a_+^{a,c_1\dots c_{k}} \right) \,,
\label{expA2}
\end{split}
\end{equation}
showing precisely how the mass eigenstates get entangled within the higher-dimensional fields. Again, this reproduces the results from \cite{Kim:1985ez}.

\paragraph{Scalars}

Let us finally sketch how to obtain the scalar mass spectrum in this example.
According to the above discussion, at level $n$ the scalar fluctuations are described
by tensoring the coset valued fluctuations (\ref{coset}) from the lowest multiplet with the 
symmetric vector representation $[n,0,0]$.
In the SO(6) basis, these fluctuations can be parametrised as
\begin{equation}
{j}_{\underline{AB},\Omega} =
\left\{
\begin{array}{l}
{j}_{ab,cd,\Omega} = 2\,\delta_{a[c} \phi_{d]b,\Omega} \,, \\
 {j}_{ab,c\alpha,\Omega} = \phi_{abcC\alpha,\Omega} \,,\\
 {j}_{a\alpha,b\beta,\Omega} = \phi_{ab,\Omega} \, \delta_{\alpha\beta} +\delta_{ab}\,\phi_{\alpha\beta,\Omega} \,,
\end{array}
\right.
\end{equation}
in terms of tensors $\phi_{ab,\Omega}$, $\phi_{\alpha\beta,\Omega}$, $\phi_{abc\alpha,\Omega}$,
constrained by
\begin{equation}
 \begin{split}
& \phi_{[ab],\Omega} = 0\,,\quad\phi_{aa,\Omega} = 0\,,\quad
\phi_{[\alpha\beta],\Omega} = 0\,,\quad\phi_{\alpha\alpha,\Omega}=0\,, \\
&
\phi_{abc\alpha,\Omega} = \varepsilon_{abcdef} \,\varepsilon_{\alpha\beta}\, \phi_{def\beta,\Omega}
\,.
\end{split}
\end{equation}
Evaluating the tensor product with the harmonics, the scalar fluctuations at level $n$
organise into the representations
\begin{equation}
 \begin{split}
\phi_{ab,c_1\dots c_n} ~&\in~
[n+2,0,0]_0 \oplus [n,0,0]_0 \oplus [n-2,0,0]_0 \oplus
[n,1,1]_0  \oplus [n-2,1,1]_0 \oplus [n-2,2,2]_0
\,,\\
\phi_{\alpha\beta,c_1\dots c_n} ~&\in~
[n,0,0]_{\pm1} \,,\\
\phi_{abc\alpha,c_1\dots c_n}  ~&\in~
[n-1,1,1]_{\pm\frac12} \oplus [n,0,2]_{+\frac12}\oplus [n,2,0]_{-\frac12} 
\oplus   [n-2,0,2]_{-\frac12}\oplus [n-2,2,0]_{+\frac12}
\,.
\end{split}
\end{equation}

From the previous results, we know that these modes still contain the unphysical
Goldstone modes
\begin{equation}
 [n-2,1,1]_0 \oplus  [n-1,1,1]_{\pm\frac12} \oplus [n,1,1]_0
\oplus [n,0,0]_{0} 
\,,
\end{equation}
of which the first three are absorbed by the massive vectors and appear with zero mass eigenvalue,
whereas the last one is absorbed into the massive spin-2 fields and must be projected out by hand.
It remains to evaluate the mass matrix \eqref{scalar_masses} on these fluctuations. The calculation is
analogous to (although somewhat more lengthy than) the ones presented above for the tensor and vector fields. 
We summarize the result for the various representations in Table~\ref{tab:masses_scalars}.

\begin{table}[tb]
\centering
\begin{tabular}{c|c|c|c}
fluctuation & representation & $m^2L_{\rm AdS}^2$ & $\Delta$
\nonumber\\
\hline
$\phi_{(\!(ab,c_1\dots c_n)\!)}$ &
$[n+2,0,0]_0$ & 
$n^2-4$ & $n+2$
\nonumber\\
$\phi_{ab,abc_1\dots c_{n-2}}$ &
$[n-2,0,0]_0$ & 
$(n+2)(n+6)$ & $n+6$
\nonumber\\
$\phi_{\alpha\beta,c_1\dots c_n}$ &
$[n,0,0]_{\pm1}$ 
& $n(n+4)$ & $n+4$
\nonumber\\
$\phi_{ab,c_1\dots c_n}$ &
$[n-2,2,2]_0$ 
& $n(n+4)$ & $n+4$
\nonumber\\
$\phi_{abc\alpha,c_1\dots c_n}$ &
$[n,0,2]_{+\frac12}\oplus [n,2,0]_{-\frac12}$ & 
$(n-1)(n+3)$ & $n+3$
\nonumber\\
$\phi_{abd\alpha,dcc_1\dots c_{n-2}}$ &
$[n-2,0,2]_{-\frac12}\oplus [n-2,2,0]_{+\frac12}$ & 
$(n+1)(n+5)$ & $n+5$
\nonumber
\end{tabular}
\caption{Masses of scalar fluctuations at level $n$\,.
The conformal dimension is given by $\Delta=2+\sqrt{4+m^2L_{\rm AdS}^2}$\,.}
\label{tab:masses_scalars}
\end{table}

\paragraph{BPS multiplets}

In the previous sections, we have determined the mass spectrum around the AdS$_5\times S^5$ background.
With the fluctuation Ansatz \eqref{eq:KKAnsatz} all mass matrices are block-diagonal level by level.
With the Ansatz \eqref{eq:KKAnsatz} for the ExFT variables, 
internal derivatives act via the combination \eqref{UUXU} acting on the twist
matrices and \eqref{defT} acting on the tower of harmonics.
The latter action is realised by the matrices \eqref{S5Tn}, such that
the resulting field equations do not mix fluctuations over different ${\rm SO}(6)$ representations $\Sigma$.
This is in contrast with the structure in the original IIB variables: after evaluating the products of the 
sphere harmonics ${\cal Y}^\Sigma$ with the twist matrices in \eqref{eq:KKAnsatz}, 
fluctuations of the original IIB fields combine linear combinations 
of different mass eigenstates as illustrated in \eqref{expB1}, \eqref{expB2} for the tensors and
in \eqref{expA1}, \eqref{expA2} for the vector fields.

The same structure underlies the ExFT supersymmetry transformations~\cite{Musaev:2014lna}. 
As a result, all fluctuations associated with a fixed ${\rm SO}(6)$ representation $\Sigma=[n,0,0]$ 
in the towers of (\ref{eq:KKAnsatz}) combine into a single $\frac12$-BPS multiplet BPS$[n]$. 
Indeed, the mass spectrum from Tables~\ref{tab:masses_tensors}--\ref{tab:masses_scalars} precisely
matches the bosonic field content of the $\frac12$-BPS multiplet BPS$[n]$ which we list in Table~\ref{tab:BPS}.
The Ansatz \eqref{eq:KKAnsatz} illustrates the fact that (except for its masses) the representation
content of the full Kaluza-Klein spectrum around a maximally symmetric vacuum such as AdS$_5\times S^5$ 
is obtained by tensoring the zero modes of the torus reduction with the tower of scalar harmonics \cite{Bianchi:2003wx}.\footnote{
This is not in contradiction with the fact that the BPS multiplet BPS$[n]$ itself does not factorize.
It is only after imposing the explicit form of the mass matrices that the degrees of freedom are distributed among the
different fields, such that for example only some of the tensor fields within the product \eqref{rep_tensors_S5_all}
actually become part of the physical spectrum.}

\begin{table}[bt]
{\scriptsize
\begin{tabular}{c|l} 
     $\Delta$ &\\ \hline
     $n+2$ & $[n+2,00]{(0\,0)}$ \\[.5ex]
     $n+\frac52$ & $[n+1,10]{(0\,\frac12)}+
[n+1,01]{(\frac12\,0)}$ \\[.5ex]
     $n+3$ & $[n,02]{(0\,0)}+[n,20]{(0\,0)}
+[n+1,00]{(0\,1)}+[n+1,00]{(1\,0)}
+[n,11]{(\frac12\,\frac12)}$ \\[.5ex]
     $n+\frac72$ & $[n,10]{(0\,\frac12)}+[n-1,12]{(0\,\frac12)}
+[n,01]{(\frac12\,0)}+[n-1,21]{(\frac12\,0)}+
[n,01]{(\frac12\,1)}+[n,10]{(1\,\frac12)}$ \\[.5ex]
     $n+4$ & $2\!\cdot\![n,00]{(0\,0)}+[n-2,22]{(0\,0)}
+[n-1,02]{(0\,1)}+[n-1,20]{(1\,0)}+2\!\cdot\![n-1,11]{(\frac12\,\frac12)}+[n,00]{(1\,1)}$
\\[.5ex]
$n+\frac92$ &
$[n-1,10]{(0\,\frac12)}+[n-2,12]{(0\,\frac12)}+
[n-1,01]{(\frac12\,0)}+[n-2,21]{(\frac12\,0)}
+[n-1,01]{(\frac12\,1)}+[n-1,10]{(1\,\frac12)}$ \\[.5ex]
     $n+5$ & $[n-2,02]{(0\,0)}+[n-2,20]{(0\,0)}
+[n-1,00]{(0\,1)}+[n-1,00]{(1\,0)}
+[n-2,11]{(\frac12\,\frac12)}$ \\[.5ex]
     $n+\frac{11}2$ & $[n-2,10]{(0\,\frac12)}+
[n-2,01]{(\frac12\,0)}$ \\[.5ex]
     $n+6$ & $[n-2,00]{(0\,0)}$ \\
\hline
\end{tabular}
}
\caption{$\frac12$-BPS multiplets of ${\rm SU}(2,2|4)$ in ${\rm SO}(6)\times{\rm SO}(4)$ notation $[n_1,n_2,n_3](j_1,j_2)$
with Dynkin labels $n_i$, and $(j_1,j_2)$ denoting the spins of ${\rm SO}(4)\sim{\rm SU}(2)\times {\rm SU}(2)$.}
\label{tab:BPS}
\end{table}

\subsubsection{${\rm AdS}_5, {\cal N}=2$, ${\rm U}(2)$ vacuum} \label{s:LeighStrassler}

In the previous section, we have worked out the Kaluza-Klein spectrum around the 
AdS$_5\times S^5$ background corresponding to the maximally symmetric stationary
point of the $D=5$ ${\rm SO}(6)$ gauged maximal supergravity of \cite{Gunaydin:1985cu}.
While this analysis reproduces the known results \cite{Kim:1985ez,Gunaydin:1984fk} for the sphere spectrum, 
our formalism allows us to address far more complicated backgrounds which are hardly accessible to standard methods.
As an illustration, let us consider another stationary point in the same scalar potential which breaks 
supersymmetry down to ${\cal N}=2$ and preserves only ${\rm SU}(2)\times {\rm U}(1)$ of the 
original ${\rm SO}(6)$ bosonic symmetry group~\cite{Khavaev:1998fb}. This stationary point can be uplifted to
a solution of IIB supergravity \cite{Pilch:2000ej}. On the field theory side of the holographic correspondence, 
this solution corresponds to the ${\cal N} = 1$ IR superconformal fixed point of the deformation of ${\cal N} = 4$ 
super-Yang-Mills by a mass term for one of the three adjoint hypermultiplets~\cite{Leigh:1995ep,Karch:1999pv}.
The holographic renormalisation group (RG) flow connecting this solution to the AdS$_5\times S^5$ background has been
constructed and studied in \cite{Freedman:1999gp}.

Within the $D=5$ supergravity of \cite{Gunaydin:1985cu}, one may compute the mass spectrum around this background for the
fields sitting within the lowest ${\cal N}=8$ multiplet which at the ${\cal N}=2$ point decomposes into
various supermultiplets of the remaining background isometry supergroup ${\rm SU}(2,2|1)\otimes{\rm SU}(2)$.
Organizing these multiplets according to their (external) ${\rm SU}(2)$ spin, this results in
 \cite{Freedman:1999gp}
\begin{equation}
\begin{split}
	{\bf [0]}~:~  &  
{\cal D}_{A_1A_1}(3;\tfrac{1}{2},\tfrac{1}{2};0) \oplus 
  {\cal D}_{LB_1}(3;0,0;+2)_{\mathbb{C}} \oplus
 {\cal D}_{LA_1}(3;0,\tfrac{1}{2};0)_{\mathbb{C}} \oplus 
 {\cal D}_{LL}(1+\sqrt{7};0,0;0)
\,, \\[2ex]
{\bf [\tfrac{1}{2}]}~:~   &  
  {\cal D}_{LB_1}(\tfrac{9}{4};\tfrac{1}{2},0;+\tfrac{3}{2})_{\mathbb{C}} 
   \oplus  {\cal D}_{LA_2}(\tfrac{11}{4};\tfrac{1}{2},0;+\tfrac{1}{2})_{\mathbb{C}} 
 \,, \\[2ex]
{\bf [1]}~:~   &  
 {\cal D}_{A_2A_2}(2;0,0;0) \oplus   {\cal D}_{LB_1}(\tfrac{3}{2};0,0;+1)_{\mathbb{C}}
 \,,
 \label{U2_level0}
\end{split}\end{equation}
where we follow the notation of~\cite{Flato:1983te} and
denote ${\rm SU}(2,2|1)$ supermultiplets by ${\cal D}(\Delta;j_1,j_2;r)$ with the arguments referring
to the conformal dimension, ${\rm SU}(2)\otimes{\rm SU}(2)$ spin and $R$-charge of the highest weight state,
respectively. 
Complex multiplets ${\cal D}(\Delta,j_1,j_2;r)_{\mathbb{C}}$
come in pairs ${\cal D}(\Delta,j_1,j_2;r)\oplus {\cal D}(\Delta,j_2,j_1;-r)$.
${\cal D}_{LL}$ denotes the generic long multiplet, while the notation for the shortening patterns 
$A_\ell$, $B_1$ for short and semi-short multiplets follows \cite{Cordova:2016emh}.

In our fluctuation Ansatz \eqref{eq:KKAnsatz} and the mass formulas worked out in
section~\ref{sec:mass}, the result (\ref{U2_level0}) corresponds to the lowest term 
in the harmonics expansion, i.e.\ to evaluating the mass matrices on the one-dimensional
space spanned by constant harmonics ${\cal Y}^{\Sigma=0}=1$, with ${\cal T}_{\underline{A}}=0$\,.
In this formalism it is then straightforward to extend the result to higher levels of the Kaluza-Klein spectrum.
As an illustration, let us give the result at level $n=1$, again with multiplets organised
according to their external ${\rm SU}(2)$ spin\footnote{
The last multiplet in the list (\ref{U2_level1}) has been missing in equation (29) of \cite{Malek:2019eaz},
where this result was first given.}
\begin{equation}
\begin{split}
	{\bf [0]}~:~  &  2 \cdot
 {\cal D}_{LA_1}(\tfrac{9}{2};0,\tfrac{1}{2};+1)_{\mathbb{C}} \oplus
     {\cal D}_{LA_1}(\tfrac{9}{2};\tfrac{1}{2},\tfrac{1}{2};+1)_{\mathbb{C}} \oplus
      {\cal D}_{LL}(\tfrac{9}{2};\tfrac{1}{2},0;+1)_{\mathbb{C}}   \\
  & \oplus
    {\cal D}_{LL}(1+\tfrac{\sqrt{37}}{2};0,0;+1) _{\mathbb{C}} \oplus
    {\cal D}_{LL}(1+\tfrac{\sqrt{61}}{2};0,0;+1)_{\mathbb{C}} \,, \\[2ex]
{\bf [\tfrac{1}{2}]}~:~   &    {\cal D}_{LB_1}(\tfrac{15}{4};0,0;+\tfrac{5}{2})_{\mathbb{C}}  \oplus
    {\cal D}_{LA_2}(\tfrac{17}{4};0,0;+\tfrac{3}{2})_{\mathbb{C}}  \oplus
     {\cal D}_{LL}(\tfrac{15}{4};\tfrac{1}{2},0;+\tfrac{1}{2})_{\mathbb{C}}   \\
  &  \oplus  {\cal D}_{LL}(\tfrac{17}{4};0,\tfrac{1}{2};+\tfrac{1}{2})_{\mathbb{C}} \oplus
   {\cal D}_{LL}(1+\tfrac{\sqrt{145}}{4};\tfrac{1}{2},\tfrac{1}{2};+\tfrac{1}{2})_{\mathbb{C}}\oplus
    {\cal D}_{LL}(1+\tfrac{\sqrt{193}}{4};0,0;+\tfrac{1}{2})_{\mathbb{C}}  \,, \\[2ex]
{\bf [1]}~:~   &   2\cdot{\cal D}_{LL}(1+\sqrt{7};0,0;0) \oplus
   {\cal D}_{LL}(1+\sqrt{7};0,\tfrac{1}{2};0)_{\mathbb{C}} \oplus
    {\cal D}_{LA_2}(\tfrac{7}{2};\tfrac{1}{2},0;+1)_{\mathbb{C}} \oplus
   {\cal D}_{LB_1}(3;\tfrac{1}{2},0;+2)_{\mathbb{C}} \;, \nonumber\\[2ex]
{\bf [\tfrac{3}{2}]}~:~   &    {\cal D}_{LB_1}(\tfrac{9}{4};0,0;+\tfrac{3}{2})_{\mathbb{C}} \oplus
    {\cal D}_{LA_2}(\tfrac{11}{4};0,0;+\tfrac{1}{2})_{\mathbb{C}}   \,.
 \label{U2_level1}
\end{split}
\end{equation}
A similar analysis can be performed at the higher Kaluza-Klein levels and be explicitly checked against 
the CFT results \cite{Bobev:2020xyz}.

\subsection{Vacua of 4-dimensional $\SO{8}$ gauged SUGRA}
We can similarly apply our mass matrices to vacua of 4-dimensional gauged supergravity, such as the $\SO{8}$-gauged SUGRA \cite{deWit:1982bul} arising from the consistent truncation of 11-dimensional supergravity on $S^7$ \cite{deWit:1984nz}. The $\SO{8}$-gauged SUGRA contains several interesting vacua from a holographic perspective. These include the maximally supersymmetric AdS$_4$ vacuum a ${\cal N}=2$ AdS$_4$ vacuum with $\SU{3} \times \mathrm{U}(1)_R$ symmetry \cite{Warner:1983du,Warner:1983vz}, and a non-supersymmetric AdS$_4$ vacuum with $\SO{3} \times \SO{3}$ symmetry \cite{Warner:1983du,Warner:1983vz}. Using the consistent truncation of 11-dimensional supergravity \cite{deWit:1984nz} all these vacua uplift to AdS solutions of 11-dimensional supergravity.

We can use our mass formulae to compute the Kaluza-Klein spectrum around these various 11-dimensional supergravity solutions. For the maximally supersymmetric AdS$_4$ vacuum, corresponding to the 11-dimensional AdS$_4 \times S^7$ solution, the Kaluza-Klein spectrum can be computed, following the steps shown in section \ref{s:S5} for the AdS$_5 \times S^5$ solution of IIB, to recover the known spectrum of AdS$_4 \times S^7$. Since the computation is analogous to that covered in detail in section \ref{s:S5}, we will not repeat it here. Instead, we will, in the following, show how our technique can be used to compute the mass spectrum of the Kaluza-Klein towers of the $\SU{3} \times \mathrm{U}(1)_R$-invariant AdS$_4$ \cite{Corrado:2001nv}, as well as the $\SO{3} \times \SO{3}$-invariant AdS$_4$ \cite{Godazgar:2014eza} vacua of 11-dimensional supergravity.

To compute the Kaluza-Klein spectra of these vacua, let us set up our notation for the $\SO{8}$ gauged supergravity. This is best described using the $\SL{8} \subset {\rm E}_{7(7)}$ subgroup under which the fundamental $\mathbf{56}$ representation of ${\rm E}_{7(7)}$ decomposes as
\begin{equation}
 \begin{split}
 \mathbf{56} &\longrightarrow \mathbf{28} \oplus \mathbf{28'} \,, \\
 \left\{ A^{\uM} \right\} &\longrightarrow \left\{ A^{ab},\, A_{ab} \right\} \,, \qquad a = 1, \ldots, 8 \,.
 \end{split}
\end{equation}
The embedding tensor of the $\SO{8}$ gauged SUGRA is given by
\begin{equation} \label{eq:SO8EmbTensor}
X_{\uM\uN}{}^{\uP}~=~\left\{ \begin{array}{l}
X_{ab,cd}{}^{ef} = - X_{ab}{}^{ef}{}_{cd} = 2\,\sqrt{2}\,\delta_{[a}^{[e}\delta^{\phantom{e}}_{b][c}\delta_{d]}^{f]} \,, \\
X_{ab}{}^{cd,ef} = 0 \,, \\
X^{ab}{}_{\uN}{}^\uP\hspace{3pt} = 0 \,. \end{array} \right.
\end{equation}

The consistent truncation of 11-dimensional SUGRA to the $\SO{8}$ gauged SUGRA can be described by a generalised Scherk-Schwarz truncation within ${\rm E}_{7(7)}$ ExFT \cite{Lee:2014mla,Hohm:2014qga}, as discussed in section \ref{subsec:genSchSch}. Just like for the consistent truncation of IIB supergravity on $S^5$, the twist matrices $U_M{}^{\uA}$ can be constructed using the elementary sphere harmonics, ${\cal Y}^a$, on $S^7$, which are just the embedding coordinates of $S^7 \subset \mathbb{R}^8$ and thus satisfy
\begin{equation}
{\cal Y}^a\, {\cal Y}^a = 1 \,.
\end{equation}
For the masses of the Kaluza-Klein spectrum, we only need to know the vector components, ${\cal K}_{\uM}$, of the corresponding generalised parallelisable frame which are given by \cite{Lee:2014mla}
\begin{equation}  \label{eq:S7vecs}
{\cal K}_{\uM} = \left\{ \begin{array}{l}
{\cal K}_{ab} = v_{ab} \,, \\
{\cal K}^{ab} = 0 \,,
\end{array} \right.
\end{equation}
where
\begin{equation} \label{eq:S7Kill}
 v_{ab}{}^m = - \sqrt{2}\, \mathring{g}^{mn} \left( \partial_{n} {\cal Y}_{[a} \right) {\cal Y}_{b]} \,,
\end{equation}
with $\mathring{g}$ the round metric on $S^7$, are the $\SO{8}$ Killing vectors of the round $S^7$.

To compute the Kaluza-Klein spectrum of any vacuum of the $\SO{8}$ gauged SUGRA, we need to choose a basis of scalar harmonics in which we expand the fields according to \eqref{eq:FlucMetric}, \eqref{eq:FlucTen} and \eqref{eq:FlucScalar}. As discussed in section \ref{s:Harm}, we can simply choose the scalar harmonics of the maximally symmetric point, which in this case is the round $S^7$. These are given, just as in section \ref{s:S5}, by the symmetric traceless polynomials in the elementary sphere harmonics ${\cal Y}^a$, i.e.
\begin{equation}
 \left\{ {\cal Y}^\Sigma \right\} = \left\{ 1,\, {\cal Y}^a,\, {\cal Y}^{ab}, \, \ldots,\, {\cal Y}^{a_1 \ldots a_n},\, \ldots \right\} \,,
\end{equation}
where ${\cal Y}^{a_1\dots a_n}\equiv {\cal Y}^{(\!(a_1} \dots {\cal Y}^{a_n)\!)}$ denotes traceless symmetrisation in the elementary harmonics. The index $\Sigma$ thus runs over the tower of symmetric vector representations $[n,0,0,0]$ of $\SO{8}$\,.

In order to evaluate the mass formulae, we need to compute the action of the vectors ${\cal K}_\uA$, defined by the generalised parallelisation \eqref{eq:S7vecs}, on the scalar harmonics ${\cal Y}^\Sigma$. For the $S^7$, these are the Killing vectors \eqref{eq:S7Kill} which, like for the $S^5$, have a linear action on the harmonics given by the generators of $\SO{8}$ in the $\left[n,0,0,0\right]$ representation,
\begin{equation}
 {\cal T}_{\uM,c_1 \ldots c_n}{}^{d_1 \ldots d_n} = n\, {\cal T}_{\uM,(\!(c_1}{}^{(\!(d_1} \delta_{c_2}^{d_2} \ldots \delta_{c_n)\!)}^{d_n)\!)} \,,
\end{equation}
in terms of the action on the elementary harmonics, given by
\begin{equation} \label{eq:SO8T}
 {\cal T}_{\uM,c}{}^{d} = \left\{ \begin{array}{l}
 {\cal T}_{ab,c}{}^{d} ~= \sqrt{2}\, \delta_{c[a} \delta_{b]}{}^d \,, \\
 {\cal T}^{ab}{}_{c}{}^{d} = 0 \,. \end{array} \right.
\end{equation}

It is now straightforward to apply our mass formulae to compute the Kaluza-Klein spectrum around any vacuum of the $\SO{8}$ gauged supergravity. All that is left to do is to dress the embedding tensor \eqref{eq:SO8EmbTensor} and the linear action on the harmonics \eqref{eq:SO8T} by the 4-dimensional scalar matrix corresponding to the vacuum of interest and apply \eqref{massSpin2}, \eqref{Mvec_gen}, \eqref{scalar_masses}.

\subsubsection{${\rm AdS}_4, {\cal N}=2$, ${\rm U}(3)$ vacuum}

We will now apply our formalism to compute the Kaluza-Klein spectrum of the 11-dimensional ${\cal N}=2$ $\SU{3} \times \mathrm{U}(1)$-invariant AdS$_4$ vacuum of 11-dimensional supergravity \cite{Corrado:2001nv}, and which can be uplifted from a vacuum \cite{Warner:1983du,Warner:1983vz} of 4-dimensional ${\cal N}=8$ $\SO{8}$ gauged SUGRA. This 11-dimensional AdS vacuum \cite{Corrado:2001nv} is similar in several respects to the AdS$_5 \times S^5$ solution dual to the Leigh-Strassler CFT discussed in section \ref{s:LeighStrassler}. The 3-dimensional ${\cal N}=2$ CFT dual is obtained by deforming the ${\cal N}=8$ ABJM CFT via a mass term for a single chiral supermultiplet and flowing to the IR. The corresponding holographic RG flow connecting the AdS$_4 \times S^7$ solution to this ${\cal N}=2$ $\SU{3} \times \mathrm{U}(1)_R$ vacuum has been constructed in \cite{Corrado:2001nv}.

Some aspects of the Kaluza-Klein spectrum of this $\SU{3} \times \mathrm{U}(1)_R$ vacuum have also already been analysed. Due to the lack of computational techniques until now, these analyses have been limited to the pattern of supermultiplets \cite{Klebanov:2008vq} and the spin-2 Kaluza-Klein spectrum \cite{Klebanov:2009kp}. Here we will use our Kaluza-Klein spectrometry to determine the full bosonic Kaluza-Klein spectrum of this 11-dimensional AdS$_4$ vacuum.

Using our mass matrices \eqref{massSpin2}, \eqref{Mvec_gen} and \eqref{scalar_masses}, we can compute the entire bosonic Kaluza-Klein spectrum of this AdS$_4$ vacuum of 11-dimensional supergravity. In fact, because the mass matrices are quadratic in $\mathrm{U}(3)$ generators and all fields organise themselves into supermultiplets, we can extrapolate the entire mass spectrum from the first three Kaluza-Klein levels alone. We find the following energies, $E_0$, for the graviton (GRAV), vector (VEC) and gravitino (GINO) supermultiplets\footnote{For the supermultiplets, we follow the notation of \cite{Klebanov:2008vq}.}
 with $\SU{3} \times \mathrm{U}(1)_R$ representation $\left[p,q\right]_r$ appearing at Kaluza-Klein level $n$:
\begin{equation}
\begin{split} \label{eq:MassU3}
\text{GRAV: } E_0 &= \frac12 + \sqrt{\frac94 + \frac12 n (n+6) - \frac43 C_{p,q} + \frac12 \left(r + \frac23(q-p) \right)^2} \,, \\
\text{GINO: } E_0 &= \frac12 + \sqrt{\frac72 + \frac12 n (n+6) - \frac43 C_{p,q} + \frac12 r^2} \,, \\
\text{VEC: } E_0 &= \frac12 + \sqrt{\frac{17}4 + \frac12 n (n+6) - \frac43 C_{p,q} + \frac12 r^2} \,,
\end{split}
\end{equation}
where $C_{p,q}$ is the eigenvalue of the representation $[p,q]$ under the quadratic Casimir operator, i.e.
\begin{equation}
C_{p,q} = \frac13 \left(p^2 + q^2 + p\,q \right) + p + q \,.
\end{equation}
Since the hypermultiplets (HYP) are necessarily short, their energies are fixed by the BPS bound but can be written similarly to the other multiplets as
\begin{equation}
E_0 = \frac12 + \sqrt{\frac{17}{4} + \frac12 n (n+6) - \frac43 C_{p,q} + \frac12 r^2} \,.
\end{equation}

The $\mathrm{U}(3)$ representations of the supermultiplets appearing at a given level $n$ can be computed by tensoring the $n = 0$ fields with the scalar harmonics and arranging these into supermultiplets. For example, the graviton supermultiplets appear at level $n$ in the representations
\begin{equation}
 \text{GRAV: } \left[p, q\right]_{\frac{p-q}3 + a - b} \,,
\end{equation}
where $p,\, q,\, a,\, b \in \mathbb{Z}^+$ are all positive integers satisfying $n = p + q + a + b$. The result for all supermultiplets appearing at levels $n \leq 3$ can be read off from the tables in \cite{Klebanov:2008vq}.

Note that the Kaluza-Klein spectrum contains infinite series of short multiplets appearing at Kaluza-Klein level $n$ with $\mathrm{U}(3)$ representation \cite{Klebanov:2008vq}
\begin{equation}
\begin{split}
\text{SGRAV: } & \left[ 0, 0 \right]_{\pm n} \,, \\
\text{SGINO: } & \left[ n+1, 0\right]_{(n+1)/3} \oplus \left[ 0, n+1 \right]_{-(n+1)/3} \,, \\
\text{SVEC: } & \left[ n+1, 1 \right]_{n/3} \oplus \left[ 1, n+1 \right]_{-n/3} \,, \\
\text{HYP: } & \left[ n+2, 0 \right]_{(n+2)/3} \oplus \left[ 0, n+2 \right]_{-(n+2)/3} \,.
\end{split}
\end{equation}
For these representations, our mass formulae \eqref{eq:MassU3} exactly reproduce the BPS bound for the short multiplets:
\begin{equation}
\begin{split}
\text{SGRAV: } E_0 &= |r| + 2 = n + 2 \,, \\
\text{SGINO: } E_0 &= |r| + \frac32 = \frac{11}6 + \frac{n}3 \,, \\
\text{SVEC: } E_0 &= |r| + 1 = \frac{n+3}{3} \,, \\
\text{HYP: } E_0 &= |r| = \frac{n+2}3 \,.
\end{split}
\end{equation}
Furthermore, our mass formulae \eqref{eq:MassU3} are valid for all supermultiplets, including long multiplets. We can thus also compute the energies of unprotected multiplets in the dual CFTs. To illustrate this, we have explicitly tabulated the energies of all multiplets appearing at levels $n \leq 2$ in tables \ref{t:U3n0} -- \ref{t:U3n2}, extending the purely group-theoretic analysis of \cite{Klebanov:2008vq} to include the energies of the supermultiplets.

\begin{table}[H]
	\begin{center}
		{\footnotesize
			\begin{tabular}{|p{27mm}|p{22mm}|p{18mm}|} \hline
				$[0,0]$ & $[0,1]$ & $[0,2]$ \\[1mm]
				MGRAV $\EOn{2}{0}$ & SGINO $\EOn{\frac{11}6}{-\frac{1}{3}}$ & HYP $\EOn{\frac23}{-\frac{2}{3}}$ \\
				LVEC $\EOn{\frac12 + \frac{\sqrt{17}}2}{0}$ & & \\[2mm] \hline
				$[1,0]$ & $[1,1]$ \\[1mm]
				SGINO $\EOn{\frac{11}6}{+\frac{1}{3}}$ & MVEC $\EOn{1}{0}$ \\[2mm] \cline{1-2}
				$[2,0]$ \\[1mm]
				HYP $\EOn{\frac23}{+\frac{2}{3}}$ \\[2mm] \cline{1-1}
			\end{tabular}
		}
		\caption{Energies of the multiplets of the ${\cal N}=2$ CFT dual to the $\mathrm{U}(3)$ AdS$_4$ vacuum at level $n=0$. We represent the energy $E_0$ and $\mathrm{U}(1)$ R-charge $r$ of a multiplet in the $[p,q]$ representation of $\SU{3}$ as $\EOn{E_0}{r}$\,.}
		\label{t:U3n0}
	\end{center}
\end{table}

\begin{table}[H]
	\begin{center}
		{\footnotesize
			\begin{tabular}{|p{35mm}|p{35mm}|p{25mm}|p{17mm}|} \hline
				$[0,0]$ & $[0,1]$ & $[0,2]$ & $[0,3]$ \\
				SGRAV $\EOn{3}{\pm1}$ & LGRAV $\EOn{\frac12 + \frac{\sqrt{145}}6}{-\frac{1}{3}}$ & SGINO $\EOn{\frac{13}6}{-\frac{2}{3}}$ & HYP $\EOn{1}{-1}$ \\
				LVEC $\EOn{\frac12 + \frac{\sqrt{3}}2}{\pm1}$ & LGINO $\EOn{\frac{17}6}{+\frac{2}{3}}$ & LVEC $\EOn{\frac73}{+\frac{1}{3}}$  &  \\[2.5mm]
				& LVEC  $\EOn{\frac12 + \frac{\sqrt{217}}6}{-\frac{1}{3}}$ &  &  \\[3mm] \hline
				$[1,0]$ & $[1,1]$ & $[1,2]$ \\
				LGRAV $\EOn{\frac12 + \frac{\sqrt{145}}6}{+\frac{1}{3}}$ & LGINO $2 \times \EOn{\frac12 + \sqrt{3}}{0}$ & SVEC $\EOn{\frac43}{-\frac{1}{3}}$ \\[2.5mm]
				LGINO $\EOn{\frac{17}6}{-\frac{2}{3}}$ &  & \\[2mm]
				LVEC  $\EOn{\frac12 + \frac{\sqrt{217}}6}{+\frac{1}{3}}$ &  &  \\[3mm] \cline{1-3}
				$[2,0]$ & $[2,1]$ \\
				SGINO $\EOn{\frac{13}6}{+\frac{2}{3}}$ & SVEC $\EOn{\frac43}{+\frac{1}{3}}$ \\
				LVEC  $\EOn{\frac73}{-\frac{1}{3}}$ &                    \\[2mm] \cline{1-2}
				$[3,0]$ \\
				HYP $\EOn{1}{+1}$ \\[1mm] \cline{1-1}
			\end{tabular}
		}
		\caption{Energies of the multiplets of the ${\cal N}=2$ CFT dual to the $\mathrm{U}(3)$ AdS$_4$ vacuum at level $n=1$. We represent the energy $E_0$ and $\mathrm{U}(1)$ R-charge $r$ of a multiplet appearing $m$ times in the $[p,q]$ representation of $\SU{3}$ as $m \times \EOn{E_0}{r}$.}
		\label{t:U3n1}
	\end{center}
\end{table}

\begin{table}[H]
	\begin{center}
		\begin{sideways}
			{\footnotesize
				\begin{tabular}{|l|l|l|l|l|} \hline
					$[0,0]$ & $[0,1]$ & $[0,2]$ & $[0,3]$ & $[0,4]$ \\
					LGRAV $\EOn{\frac12 + \frac{\sqrt{41}}2}{0}$ & conj. to $[1,0]$ & conj. to $[2,0]$ & conj. to $[3,0]$ & conj. to $[4,0]$ \\[1.5mm]
					SGRAV $\EOn{4}{\pm2}$ &  &  &  & \\[1mm]
					LVEC $\EOn{\frac12 + \frac{\sqrt{57}}2}{\pm2}$, $2 \times \EOn{4}{0}$ &  &  & & \\ \hline
					$[1,0]$ & $[1,1]$ & $[1,2]$ & $[1,3]$ \\
					LGRAV $\EOn{\frac12 + \frac{\sqrt{337}}6}{-\frac{2}{3}}$, $\EOn{\frac12 + \frac{\sqrt{313}}6}{+\frac{4}{3}}$ & LGRAV $\EOn{3}{0}$ & conj. to $[2,1]$ & conj. to $[3,1]$ \\[3mm]
					LGINO $\EOn{\frac{23}{6}}{-\frac{5}{3}}$, $2 \times \EOn{\frac12 + \frac{2\sqrt{22}}3}{+\frac{1}{3}}$ & LGINO $2 \times \EOn{\frac12 + 2\sqrt{2}}{\pm1}$ &  & \\[3mm]
					LVEC $\EOn{\frac12 + \frac{\sqrt{385}}6}{-\frac{2}{3}}$, $\EOn{\frac12 + \frac{\sqrt{409}}6}{+\frac{4}{3}}$ & LVEC $2 \times \EOn{\frac12 + \frac{\sqrt{33}}2}{0}$ & & \\ \cline{1-4}
					$[2,0]$ & $[2,1]$ & $[2,2]$ \\
					LGRAV $\EOn{\frac12 + \frac{\sqrt{217}}6}{+\frac{2}{3}}$ & LGINO $2 \times \EOn{\frac12 + \frac{2\sqrt{10}}3}{+\frac13}$ & LVEC $\EOn{\frac12 + \frac12 \sqrt{\frac{19}{3}}}{0}$ \\
					LGINO $\EOn{\frac{19}6}{-\frac{1}{3}}$ & LVEC $\EOn{\frac12 + \frac{\sqrt{193}}6}{-\frac{2}{3}}$ & \\
					LVEC $\EOn{\frac12 + \frac{\sqrt{313}}6}{-\frac{4}{3}}$, $2 \times \EOn{\frac{10}3}{+\frac23}$ & & \\ \cline{1-3}
					$[3,0]$ & $[3,1]$ \\[1mm]
					SGINO $\EOn{\frac{15}{6}}{+1}$ & SVEC $\EOn{\frac53}{+\frac{2}{3}}$ \\
					LVEC $\EOn{\frac12+\frac12\sqrt{\frac{19}3}}{0}$   & \\ \cline{1-2}
					$[4,0]$ \\[0.5mm]
					HYP $\EOn{\frac43}{+\frac{4}{3}}$ \\ \cline{1-1}
				\end{tabular}
			}
		\end{sideways}
		\caption{Energies of the multiplets of the ${\cal N}=2$ CFT dual to the $\mathrm{U}(3)$ AdS$_4$ vacuum at level $n=2$. We represent the energy $E_0$ and $\mathrm{U}(1)$ R-charge $r$ of a multiplet appearing $m$ times in the $[p,q]$ representation of $\SU{3}$ as $m \times \EOn{E_0}{r}$\,.}
		\label{t:U3n2}
	\end{center}
\end{table}

\subsubsection{${\rm AdS}_4, {\cal N}=0$, ${\rm SO}(4)$ vacuum}
The $\SO{8}$ gauged SUGRA also contains a prominent non-supersymmetric AdS$_4$ vacuum with $\SO{3} \times \SO{3}$ symmetry \cite{Warner:1983du,Warner:1983vz}, whose uplift to 11-dimensional supergravity was constructed in \cite{Godazgar:2014eza}. Intriguingly, this vacuum is stable within the ${\cal N}=8$ 4-dimensional supergravity, with all scalar fields above the Breitenlohner-Freedman (BF) bound. It was long hoped that the AdS$_4$ vacuum would also be stable within 11-dimensional supergravity, but since the AdS$_4$ vacuum is not supersymmetric and has few symmetries, computing its Kaluza-Klein spectrum has remained elusive.

However, using the technique laid out here, we can exploit the fact that this AdS$_4$ vacuum arises by deforming AdS$_4 \times S^7$ by modes living within the $\SO{8}$ consistent truncation. As a result, it is straightforward to compute the bosonic Kaluza-Klein spectrum using our mass formulae \eqref{massSpin2}, \eqref{Mvec_gen} and \eqref{scalar_masses}, as was done in \cite{Malek:2020mlk} up to level 6 above the 4-dimensional ${\cal N}=8$ supergravity. The Kaluza-Klein spectrum displays the curious feature that the masses of the Kaluza-Klein modes does not increase monotonically with the level $n$. Instead, even though the Kaluza-Klein scalars at levels 0 and 1 are stable, the Kaluza-Klein tower contains tachyonic scalar fields at levels 2 and higher whose masses violate the BF bound.
Therefore, the techniques developed here show that this non-supersymmetric AdS$_4$ vacuum is unstable within 11-dimensional supergravity, lending further evidence to the Swampland Conjecture \cite{Ooguri:2016pdq} that all non-supersymmetric AdS vacua of string theory must be unstable.

Specifically, the scalar mass matrix \eqref{scalar_masses} at level 0 yields the following mass eigenvalues
\begin{equation}
\begin{split}
	{\bf (0,0)}   ~:~  &  
\{
-1.714\,(2)\,,8.571
 \}
\,, \\[1ex]
{\bf (1,1)}    ~:~  & 
\{
-1.714\, (2)\,, -1.312, 2.571, 5.598 \}
  \,,
 \label{SO4_level0}
\end{split}
\end{equation}
normalised in units of the inverse AdS length square, 
where ${\bf (j_1,j_2)}$ denotes the ${\rm SO}(4)\sim {\rm SU}(2)\otimes{\rm SU}(2)$ representations\footnote{
In \cite{Malek:2020mlk}, we have used the notation ${\bf (2j_1+1,2j_2+1)}$ for these representations.
}, and where the states with mass $m^2 L_{\rm AdS}^2=-1.714$ appear with multiplicity 2\,.
This reproduces the result of \cite{Fischbacher:2010ec} and shows that within ${\cal N}=8$ supergravity, all scalar masses lie 
above the BF bound  $m_{\rm BF}^2 L_{\rm AdS}^2=-2.25$\,.

Evaluating the mass matrix at level $n=1$, we obtain the masses
\begin{equation}
\begin{split}
	{\bf (\tfrac12,\tfrac12)}   ~:~  &  
\{
-2.232, -2.225, -1.947, -0.752, 3.790, 5.059, 5.766, 7.627, \\
&  ~~ 10.567, 
10.707, 11.492, 16.004 \}
\,, \\[1ex]
{\bf (\tfrac12,\tfrac32)} \oplus {\bf (\tfrac32,\tfrac12)}    ~:~  & 
\{ -1.196, -0.996, 1.732, 2.429, 6.198, 6.292, 9.817, 11.725 \}
 \,, \\[1ex]
{\bf (\tfrac32,\tfrac32)}    ~:~  & 
 \{ -1.965, -1.377, -0.761, 1.042, 3.208, 3.431, 3.831, 7.497, 7.882, 
12.999 \}
 \,,
 \label{SO4_level1}
\end{split}
\end{equation}
still all lying above the BF bound. However, at level $n=2$, the mass eigenvalues
are given by
\begin{equation}
\begin{split}
	{\bf (0,0)}   ~:~  &  
\{
-3.117, -2.821, -2.179, 0.941, 1.995, 3.181, 5.244, 6.753, 7.224, 
9.838,  \\
&  ~~ 12.000 , 12.108, 12.221, 14.764, 16.685, 18.000, 19.418, 19.613, 
24.702 \}
\,, \\[1ex]
{\bf (1,1)}    ~:~  & 
\{-2.532, -2.448, -1.220, 0. \,(3)\,, 0.846, 1.483, 2.586, 2.884, 4.133, 4.228, 
4.400, 
 \\
&  ~~5.239, 6.282, 6.450, 6.964, 7.613, 7.793, 9.017, 9.685, 9.806, 
10.002, 11.456, 11.462, 
 \\
&  ~~12.196, 12.767, 12.871, 13.010, 14.066, 
14.556, 15.257, 18.839, 19.385, 20.107, 26.532 \}
 \,, \\[1ex]
{\bf (2,2)}    ~:~  & 
 \{ -2.361, -0.916, 0.224, 2.291, 4.212, 4.419, 5.467, 6.513, 9.429, 
10.286, 10.980, 
 \\
&  ~~11.822, 15.295, 16.464, 23.305 \}
 \,, \\[1ex]
{\bf (0,1)}\oplus {\bf (1,0)}    ~:~  & 
 \{-1.343, -0.232, 3.050, 3.725, 5.697, 6.731, 8.032, 9.647, 10.087, 
11.597, 12.510, 
 \\
&  ~~13.574, 14.955, 15.952, 17.676, 21.337, 21.862 \}
 \,, \\[1ex]
{\bf (0,2)}\oplus {\bf (2,0)}     ~:~  & 
 \{ -0.975, -0.110, 2.410, 3.175, 5.301, 7.183, 7.588, 9.731, 11.241, 
12.232, 14.261, 
 \\
&  ~~16.019, 18.407, 23.822 \}
 \,, \\[1ex]
{\bf (1,2)}\oplus {\bf (2,1)}   ~:~  & 
 \{-0.881, -0.203, 2.143, 3.161, 3.245, 4.430, 4.984, 7.480, 7.946, 8.592, 
9.234, 
 \\
&  ~~ 12.032, 12.855, 13.948, 14.334, 18.746, 21.097 \}
 \,,
 \label{SO4_level2}
\end{split}
\end{equation}
and include a number of tachyonic modes $m^2 L_{\rm AdS}^2<-2.25$\,.
Similarly, tachyonic modes are found at the higher Kaluza-Klein levels~\cite{Malek:2020mlk} .

Moreover, the result \eqref{SO4_level2} for the Kaluza-Klein modes at level 2 shows 
27 physical massless scalar fields (i.e.\ massless scalars not eaten by massive vector or graviton fields), which transform in the $3 \cdot \mathbf{\left(1,1\right)}$ of the $\SO{3} \times \SO{3}$ symmetry group. These scalars are thus infinitesimal moduli which break the $\SO{3} \times \SO{3}$ symmetry. If these AdS$_4$-preserving deformations can be integrated up to finite moduli, then this would give rise to a family of non-supersymmetric AdS$_4$ vacua of 11-dimensional supergravity with symmetries smaller than $\SO{3} \times \SO{3}$.

\section{Conclusions} \label{s:Conclusions}

In this paper, we have shown how the formalism of exceptional field theory can be used as a powerful tool for the computation of the complete Kaluza-Klein mass spectra around vacua that lie within consistent truncations. In particular, the method applies to deformed backgrounds that may have little or no isometries left, as well as to non-supersymmetric backgrounds. We have derived the explicit form of the mass matrices \eqref{massSpin2}, \eqref{massB}, \eqref{Mvec_gen}, \eqref{scalar_masses}, for compactifications to $D=4$ and $D=5$ dimensions, that are described within E$_{7(7)}$ and E$_{6(6)}$ ExFT, respectively. They are given in terms of the embedding tensor characterizing the consistent truncation to the lowest multiplet, together with the (dressed) action on the scalar harmonics associated with the maximally symmetric point within this consistent truncation. In terms of the ExFT variables, the fluctuations are described by a simple product Ansatz \eqref{eq:KKAnsatz} between the Scherk-Schwarz twist matrices and the tower of scalar fluctuations. Translating this back into the original supergravity variables allows us to straightforwardly identify the resulting mass eigenstates in higher dimensions.

We have illustrated the formalism in various examples. First, we have re-derived the full bosonic Kaluza-Klein spectrum around the maximal symmetric AdS$_5\times S^5$ solution of IIB supergravity, finding agreement with the classic results of \cite{Kim:1985ez,Gunaydin:1984fk}. Next, we have applied the method to compute the higher Kaluza-Klein levels around some prominent AdS vacua with less supersymmetry in $D=5$ and $D=4$ dimensions. This provides valuable information for various holographic dualities and for the stability analysis of non-supersymmetric vacua. Although in this paper, we have restricted the analysis to the bosonic mass spectrum, it is clear that the fermionic mass spectrum can be computed in complete analogy based on the structures of supersymmetric ExFT \cite{Godazgar:2014nqa,Musaev:2014lna}. Also, while we have restricted our examples to AdS vacua which are of particular interest in the holographic context, the method and the explicit mass matrices likewise apply for Minkowski and dS vacua.

There are many further potential applications of the methods presented in this paper. Some recent and rather exhaustive scans of the potentials of maximal $\SO{8}$ gauged supergravity in $D=4$ \cite{Comsa:2019rcz} and $\SO{6}$ gauged supergravity in $D=5$ \cite{Krishnan:2020sfg,Bobev:2020ttg} have revealed a plethora of AdS vacua, most of which preserve very few bosonic (and super-)symmetries. Our analysis of the Kaluza-Klein spectrum can be applied to all of these. Likewise, our method applies to vacua within other maximal supergravities, such as the $D=4$, $\mathrm{ISO}(7)$ gauged supergravity which describes the consistent truncation of massive IIA supergravity \cite{Guarino:2015jca} on $S^6$ and exhibits a rich vacuum structure \cite{Guarino:2015qaa}. In this case, the maximally symmetric point, which is used to construct the scalar harmonics, would be the round $S^6$, even though this is not a vacuum of the theory. Another interesting gauging is the $D=4$ SUGRA with $[{\rm SO}(1, 1) \times {\rm SO}(6)] \ltimes \mathbb{R}^{12}$ gauge group whose potential carries numerous interesting AdS vacua \cite{DallAgata:2011aa,Guarino:2020gfe} with IIB origin~\cite{Inverso:2016eet}. The analysis of their Kaluza-Klein spectra will require a proper treatment of the non-compact gauge group generator whose associated non-compact direction will have to undergo a proper S-folding in order to extract a discrete spectrum of harmonics. For the spin-2 spectrum, this was analysed, for example, in \cite{Dimmitt:2019qla}.

We have derived in this paper the explicit mass matrices for E$_{7(7)}$ and E$_{6(6)}$ ExFT. However, the fluctuation Ansatz 
\eqref{eq:KKAnsatz} is universal and allows to work out the mass matrices for other exceptional field theories, giving
rise to the Kaluza-Klein spectra in compactifications to other dimensions.
It would also be very interesting to extend the formalism to vacua sitting in consistent truncations that preserve a lower number
of supersymmetries building on the constructions of \cite{Malek:2017njj,Cassani:2019vcl}.

\paragraph*{Acknowledgements}

We would like to thank
D. Andriot,
N. Bobev,
C. Eloy, 
A. Guarino,
M. Gutperle,
O. Hohm,
G.~Larios,
H. Nicolai,
C. Nunez,
B. Robinson,
E. Sezgin,
A. Tomasiello, 
J. van Muiden,
O. Varela,
D. Waldram,
and
N. Warner
for useful discussions and comments.
We acknowledge the Mainz Institute for Theoretical Physics (MITP) of the Cluster of Excellence PRISMA+ (Project ID 39083149) for hospitality while this work was initiated. EM is supported by the Deutsche Forschungsgemeinschaft (DFG, German Research Foundation) via the Emmy Noether program ``Exploring the landscape of string theory flux vacua using exceptional field theory'' (project number 426510644).


\providecommand{\href}[2]{#2}\begingroup\raggedright\endgroup

\end{document}